\documentclass[a4paper]{article}

\usepackage{epsfig}
\usepackage{subfigure}
\usepackage{amssymb}
\usepackage{amsmath}
\usepackage{amssymb}
\usepackage{tabularx}
\usepackage{wrapfig}
\usepackage{algorithm}
\usepackage{algorithmic}
\usepackage{booktabs}
\usepackage[plainpages=false]{hyperref}
\usepackage{xspace}
\usepackage[english]{babel}
\usepackage{amsopn}
\usepackage{latexsym}
\usepackage{textcomp}
\usepackage{color, colortbl}
\usepackage{times}
\usepackage{enumerate}
\usepackage{tabularx}
\usepackage[figure,table]{hypcap}
\usepackage{multirow}
\usepackage{todonotes}

\newcommand{\remove}[1]{}
\newcommand{\spinner}{{\tt Spinner}\xspace}

\newcommand{\algo}{\texttt{GiLA}\xspace}

\begin{document}



\title{A Distributed Force-Directed Algorithm on Giraph: Design and Experiments\thanks{Research supported in part by the MIUR project AMANDA ``Algorithmics for MAssive and Networked DAta'', prot. 2012C4E3KT\_001. A preliminary extended abstract of this paper has been accepted to the 23th International Symposium on Graph Drawing and Network Visualization (GD'15).}}

\author{Alessio Arleo\thanks{Universit{\`a} degli Studi di Perugia, Italy, \texttt{\{name.surname\}@unipg.it}} \and Walter Didimo\footnotemark[2] \and \and Giuseppe Liotta\footnotemark[2]  \and Fabrizio Montecchiani\footnotemark[2]}

\date{}

\maketitle

\begin{abstract}
In this paper we study the problem of designing a distributed graph visualization algorithm for large graphs. The algorithm must be simple to implement and the computing infrastructure must not require major hardware or software investments. We design, implement, and experiment a force-directed algorithm in Giraph, a popular open source framework for distributed computing, based on a vertex-centric design paradigm. The algorithm is tested both on real and artificial graphs with up to million edges, by using a rather inexpensive PaaS (Platform as a Service) infrastructure of Amazon. The experiments show the scalability and effectiveness of our technique when compared to a centralized implementation of the same force-directed model. We show that graphs with about one million edges can be drawn in less than 8 minutes, by spending about 1\$ per drawing in the cloud computing infrastructure. 
\end{abstract}

\section{Introduction}\label{se:introduction}
The automatic visualization of graphs is a central activity for analyzing and mining relational data sets, collected and managed through the different kinds of information systems and information science technologies. Examples occur in many applications domains, including social sciences, computational biology, software engineering, Web computing, information and homeland security (see, e.g.,~\cite{dett-gd-99,dl-gvdm-07,jm-gds-03,kw-dg-01,s-gd-02,t-hgd-13}). 

Classical force-directed algorithms, like \emph{spring embedders}, are by far the most popular graph visualization techniques (see, e.g.,~\cite{dett-gd-99,Handbook-kob}). One of the key components of this success is the simplicity of their implementation and the effectiveness of the resulting drawings. Spring embedders and their variants make the final user only a few lines of code away from an effective layout of a network. They model the graph as a physical system, where vertices are equally-charged electrical particles that repeal each other and edges act like springs that give rise to attractive forces. Computing a drawing corresponds to finding an equilibrium state of the force system by a simple iterative approach (see, e.g.,~\cite{e-hgd-84,fr-gdfdp-91}).

The main drawback of spring embedders is that they are relatively expensive in terms of computational resources, which gives rise to scalability problems even for graphs with a few thousands vertices. To overcome this limit, sophisticated variants of force-directed algorithms have been proposed; they include \emph{hierarchical space partitioning}, \emph{multidimensional scaling}, \emph{stress-majorization}, and \emph{multi-level} techniques (see, e.g.,~\cite{DBLP:conf/gd/BartelGKM10,DBLP:journals/ivs/GibsonFV13,DBLP:conf/gd/HachulJ04,Handbook-kob} for surveys and experimental works on these approaches). Also, both centralized and parallel multi-level force-directed algorithms that use the power of graphical processing units (GPU) have been designed and implemented~\cite{DBLP:conf/gd/GodiyalHGH08,DBLP:journals/tvcg/IngramMO09,skssl-sunc+-2011,yya-sffg+-12}. They scale to graphs with some million edges, but their development requires a low-level implementation and the necessary infrastructure is typically expensive in terms of hardware and maintenance.

\paragraph{Our Contributions} Motivated by the increasing availability of scalable cloud computing services, we study the problem of designing a simple force-directed algorithm for a distributed architecture. We want to use such an algorithm on an inexpensive PaaS (Platform as a Service) infrastructure to compute drawings of graphs with million edges. Our contributions are as follows:

\begin{itemize}
\item We give a new distributed force-directed algorithm, based on the Fruchterman-Reingold model~\cite{fr-gdfdp-91}, designed according to the ``{\em Think-Like-A-Vertex (TLAV)}'' paradigm. TLAV is a vertex-centric approach to design a distributed algorithm from the perspective of a vertex rather than of the whole graph. It improves locality, demonstrates linear scalability, and can be easily adopted to reinterpret many centralized iterative graph algorithms~\cite{mwm-tlvs+-2015}. Also, it overcomes the limits of other popular distributed paradigms like MapReduce, which are often poor-performing for iterative graph algorithms~\cite{mwm-tlvs+-2015}.  

\item We describe an implementation and engineering of our algorithm within the \emph{Apache Giraph} framework~\cite{c-glgp+-2011}, a popular open-source platform for TLAV distributed graph algorithms. For example, Giraph is used by Facebook to efficiently analyze the huge network of its users and their connections~\cite{DBLP:journals/pvldb/ChingEKLM15}. The code of our implementation is made available over the Web ({\small\url{http://www.geeksykings.eu/gila/}}), to be easily re-used in further research on distributed graph visualization algorithms.   

\item We present the results of an extensive experimental analysis of our algorithm on a small Amazon cluster of up to 21 computers, each equipped with 4 vCPUs ({\small\url{http://aws.amazon.com/en/elasticmapreduce/}}). The experiments are performed both on real and artificial graphs, and show the scalability and effectiveness of our technique when compared to a centralized version of the same force-directed model. The experimental data also show the very limited cost of our approach in terms of cloud infrastructure usage. For example, computing a drawing on a set of graphs of our test suite with one million edges requires on average less than 8 minutes, which corresponds to about 1\$ (USD) payed to Amazon.   

\item Finally, we describe an application of our drawing algorithm to visual cluster detection on large graphs. The algorithm is easily adapted to compute a layout of the input graph using the \emph{LinLog} force model proposed by Noack~\cite{DBLP:journals/jgaa/Noack07}, which is conceived to geometrically emphasize clusters. On this layout, we define and highlight clusters of vertices by using the $K$-means algorithm, and experimentally evaluate the quality of the computed clustering.
\end{itemize}

\paragraph{Structure of the paper} The remainder of the paper is structured as follows. 
Section~\ref{se:related} discussed further work related to our research.
Section~\ref{se:bg} provides the necessary background on the force-directed model adopted in our solution and on the TLAV paradigm within Giraph. In Section~\ref{se:vcse} we describe in details our distributed algorithm, the related design challenges, and its theoretical computational complexity. In Section~\ref{se:experiments} we present our implementation and the results of our experimental analysis. Section~\ref{se:application} shows the application of our algorithm to visual cluster detection on large graphs. Finally, in Section~\ref{se:future} we discuss future research directions for our work.

\section{Related Work}\label{se:related}

So far, the design of distributed graph visualization algorithms has received limited attention.  

Mueller \emph{et al.}~\cite{DBLP:conf/egpgv/MuellerGL06} and Chae \emph{et al.}~\cite{csmag-hdgv+-2012} proposed force-directed algorithms that use multiple large displays. Vertices are evenly distributed on the different displays, each associated with a different processor, which is responsible for computing the positions of its vertices; scalability experiments are limited to graphs with some thousand vertices. Tikhonova and Ma~\cite{DBLP:conf/egpgv/TikhonovaM08} presented a parallel force-directed algorithm that can run on graphs with few hundred thousand edges. It takes about 40 minutes for a graph of 260,385 edges, on 32 processors of the PSC's BigBen Cray XT3 cluster. 

The algorithms mentioned above are mainly \emph{parallel} algorithms, rather than \emph{distributed} algorithms. Their basic idea is to partition the set of vertices among the processors and keep data locality as much as possible throughout the computation. 

More recently, Hinge and Auber~\cite{DBLP:conf/iv/AntoineD15} described a distributed force-directed algorithm implemented in the Spark framework ({\small\url{http://spark.apache.org/}}), using the graph processing library GraphX. Their approach is mostly based on a MapReduce paradigm instead of a TLAV paradigm. As for our work, the research of Hinge and Auber goes in the direction of exploring emerging frameworks for distributed graph algorithms. However, the approach shows margins for improvement: their algorithm takes 5 hours to compute the layout of a graph with just 8,000 vertices and 35,000 edges, on a cluster of 16 machines, each equipped with 24 cores and 48GB of RAM.     

\section{Background}\label{se:bg}

We first recall the Fruchterman-Reingold force-directed model (Section~\ref{sse:fr-alg}), which is central for the description of our distributed algorithm. Afterwards, we recall the basic concepts behind the TLAV paradigm and the Giraph framework (Section~\ref{sse:giraph}). 

\subsection{Fruchterman-Reingold Force-Directed Algorithm}\label{sse:fr-alg}
Two main simple principles have guided the design of several force-directed algorithms over the years (see, e.g.,~\cite{e-hgd-84,fr-gdfdp-91}): $(i)$ vertices connected by an edge should be drawn near to each other; $(ii)$ vertices should not be drawn too close to each other. We point the reader to the comprehensive survey by Kobourov~\cite{Handbook-kob} for references and explanations on the wide literature on this subject. The common ingredients of force-directed algorithms are a model of the physical system of forces acting on the vertices and an iterative algorithm to find a static equilibrium of this system. 
Here, we restrict our attention to the Fruchterman-Reingold (FR for short) force-directed algorithm~\cite{fr-gdfdp-91}. 

According to the FR model, vertices can be viewed as equally-charged electrical particles and edges act similar to springs; the electrical charges cause repulsion between vertices, while the springs cause attraction. Thus, only vertices that are neighbors attract each other, but all vertices repel each other. Starting from an initial configuration (which usually corresponds to a random placement), the FR algorithm executes a suitable number of iterations, seeking for a static equilibrium of the above defined mechanical system. In each iteration, the FR algorithm computes the effect of attractive and repulsive forces on each vertex. The total displacement of a vertex is limited to some maximum value, which decreases over the iterations. Such a maximum value is often called the \emph{temperature} of the system, since its decreasing has the effect of cooling down the system. Let $G=(V,E)$ be the input graph; the resulting force acting on a vertex $u \in V$ during each iteration is as follows:

$$F(u) = \sum_{(u,v)\in E}f^a_{uv}(u)+\sum_{(u,v) \in V \times V}f^r_{uv}(u),$$

where $f^a_{uv}$ and $f^r_{uv}$ denote the attractive force and the repulsive force exerted by $v$ on $u$, respectively. More in detail, denoted by $\delta(u,v)$ the Euclidean distance between vertices $u$ and $v$, the absolute values of the forces on $u$ are as follows:

$$\Vert f^a_{uv} \Vert = \frac{\delta^p(u,v)}{d}\;\;\;\;\;\;\;\;\;\;\;\;\;\Vert f^r_{uv} \Vert = \frac{d^2}{\delta^q(u,v)}$$

In the above formulas, $d$ is a constant that represents the ideal distance between two vertices, while $p$ and $q$ are two integers that can be suitably tuned based on the structure of the input graphs and on the particular application for which the drawings are computed. Clearly, this model is only an approximation of a realistic physical model.
In~\cite{fr-gdfdp-91} they are chosen such that $p=2$ and $q=1$ (see also Section~\ref{sse:implementation}). However, other variants have been proposed such as the one by Noack~\cite{DBLP:journals/jgaa/Noack07} (see Section~\ref{se:application}) .  

\subsection{The TLAV paradigm and the Giraph framework}\label{sse:giraph}
The vertex-centric programming model is a paradigm for distributed processing frameworks to address computing challenges on large graphs. The idea behind this model is to ``Think-Like-A-Vertex'' (TLAV), which translates to implementing distributed algorithms from the perspective of a vertex rather than of the whole graph. Each vertex can store a limited amount of data and can exchange messages only with its neighbors. TLAV frameworks provide a common vertex-centric programming interface, abstracting from low-level details of distributed computation, thus improving re-usability and maintainability of program source codes. 

Google's Pregel~\cite{mabd+-pslgp-2010}, based on the Bulk Synchronous Programming model (BSP)~\cite{v-bmpc-1990}, was the first published implementation of a TLAV framework. \emph{Giraph}~\cite{c-glgp+-2011} is a popular TLAV framework built on the Apache Hadoop infrastructure and originated as the open source counterpart of Pregel. Giraph adds several features to the basic Pregel model, although it is still based on the BSP model. In Giraph, the computation is split into \emph{supersteps} executed iteratively and synchronously. A superstep consists of two processing steps: 

\begin{enumerate}
\item Each vertex executes a user-defined vertex function based on both local vertex data and on data coming from its adjacent vertices;

\item Each vertex sends the results of its local computation to its neighbors, along its incident edges.
\end{enumerate}

The whole computation ends once a fixed number of supersteps has occurred or when certain user-defined conditions are met (i.e., no message has been sent or an equilibrium state is reached). Giraph also allows the definition of global variables (called \emph{aggregators}), which can be accessed by each vertex during the computation. However, aggregators should be used carefully as they undermine the basic principles of TLAV and may result in an expensive synchronization cost.

\section{A Vertex-Centric Distributed Force-Directed Algorithm}\label{se:vcse}

We first discuss the main challenges that must be addressed in the design a force-directed algorithm for a TLAV framework such as Giraph (Section~\ref{sse:challenges}). These challenges are mainly related to the calculation of the repulsive forces acting between each pair of vertices, which requires each vertex to know the positions of all the other vertices. After this discussion, we describe our algorithmic solution (Sections~\ref{sse:overview}--\ref{sse:complexity}), based on a pipeline involving several steps of computation. 

\subsection{Design Challenges}\label{sse:challenges}
Sequential, shared-memory graph algorithms are inherently centralized. They usually receive the entire graph as input, presume all data is accessible in memory, and the graph is processed in a sequential manner. As already observed, a more local and distributed approach is more suitable (and often necessary) for processing very large graphs. Hence, a paradigm shift from centralized to decentralized approaches is needed and, depending on the algorithm, this shift may give rise to some critical challenges.

In the following we discuss the main challenges to be addressed when designing a force-directed algorithm for a TLAV paradigm. For the sake of presentation, we focus on the FR force-directed algorithm and on the Giraph TLAV framework, although most of these considerations hold more in general. The following three properties must be guaranteed in the design of a TLAV-based algorithm:

\begin{itemize}

\item[{\tt P1.}] Each vertex exchange messages only with its neighbors;

\item[{\tt P2.}] Each vertex locally stores a small amount of data;

\item[{\tt P3.}] The communication load in each supertsep (i.e., the total number and length of messages sent in the superstep) is small: for example, linear in the number of edges of the graph.
\end{itemize} 

Property {\tt P1} usually corresponds to an architectural constraint of the TLAV framework. Violating {\tt P2} causes out-of-memory errors during the computation of large instances. Violating {\tt P3} quickly leads to inefficient computations, especially on graphs that are locally dense or that have high-degree vertices.   

Each iteration of the FR algorithm can be directly mapped onto a set of consecutive supersteps of the computation. Within each iteration, we need to compute the forces acting on each vertex, which are then used to update the corresponding positions. Recall that, while attractive forces only act on pairs of adjacent vertices, repulsive forces act on every pair of vertices. The above three constraints {\tt P1}--{\tt P3} do not allow for simple strategies to make a vertex aware of the positions of all other vertices in the graph, which is required to compute the repulsive forces acting on the vertex. For instance, a possible solution that respects {\tt P1} and {\tt P2} is to propagate the position of every vertex throughout the network using a flooding technique. However, such a strategy would generate an unmanageable communication load, thus violating {\tt P3}. On the other hand, a more sophisticated routing protocol would require large routing tables stored at each vertex, which collides with {\tt P2}. We address these challenges by adopting a locality principle, as explained in Section~\ref{sse:layout}. 

Moreover, since computing the repulsive forces between each pair of vertices has a quadratic cost in a centralized paradigm, some force-directed algorithms use spatial decomposition to efficiently approximate forces between pair of vertices that are far from each other (see, e.g.,~\cite{DBLP:conf/gd/HachulJ04,DBLP:conf/gd/QuigleyE00}). These approaches still require the knowledge of the whole graph and thus present analogous issues for a decentralized algorithm as those already discussed.

\subsection{Algorithmic Pipeline}\label{sse:overview}
\begin{figure}[t]
\centering
\includegraphics[width=1\columnwidth]{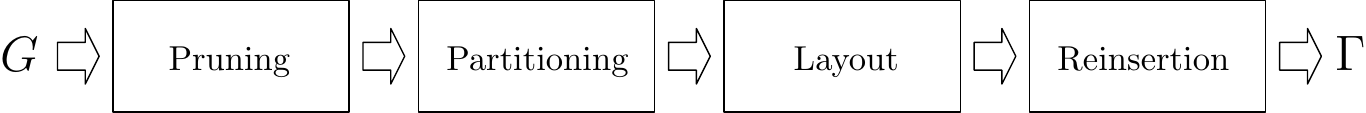}
\caption{\small Algorithm pipeline of \algo: $G$ is the input graph and $\Gamma$ the computed drawing.}\label{fi:pipeline}
\end{figure}

Our distributed force-directed algorithm, called \algo, is simple and is designed to run on a cluster of computers within a TLAV framework. It adopts the same model as the FR algorithm and consists of the algorithmic pipeline depicted in Figure~\ref{fi:pipeline}. Each step of this pipeline is described in detail in Subsections~\ref{sse:pruning}--\ref{sse:resinsertion}. The computational complexity of this pipeline is analyzed in Subsection~\ref{sse:complexity}.

\subsection{Pruning}\label{sse:pruning}
In this first step, the algorithm handles vertices that have only one neighbor (i.e., degree-one vertices). 
%Indeed, we experimentally observed that real-world networks, as well as synthetic complex networks, often contain a significant amount of vertices of degree one (see also Section~\ref{sse:exp-setting})\footnote{We should report the percentages of degree-one vertices in our tables}. 
For the sake of efficiency, we initially remove such vertices; they will be reinserted at the end of the computation with an ad-hoc technique. This operation is performed right after the graph has been loaded in memory. The number of degree-one vertices adjacent to a vertex $v$ is stored as a local information of $v$, to be used throughout the computation. 

\subsection{Partitioning}\label{sse:partitioning}
Large-scale graphs must be divided into parts to be placed in a distributed memory.  Good partitions often lead to improved performance, but expensive strategies may end up dominating the processing time. Effective partitioning evenly distribute the vertices for balanced workload, while minimizing inter-partition edges to avoid costly network traffic. A well-known partitioning algorithm is METIS~\cite{DBLP:conf/icpp/KarypisK95}. See also~\cite{mwm-tlvs+-2015} for further details and references. 

In Giraph a computing unit is called \emph{worker}, and each computer can host more than one worker. Giraph provides a hash-based algorithm to assign each vertex of the graph to a particular worker. Although the Giraph default partitioning algorithm is very fast and provides balanced partitions, it may not produce good results in terms of locality (i.e., minimization of the edges between vertices assigned to different units).  Recently, Vaquero {\em et al.}~\cite{DBLP:conf/icdcs/VaqueroCLM14} introduced \spinner, a partitioning algorithm for Giraph based on iterative vertex migrations, relying only on local information. Starting from a random initial label assignment, which corresponds to an initial partitioning, vertices iteratively adopt the label of the majority of their neighbors  until no new labels are assigned (i.e., convergence is reached). On every iteration, each vertex $v$ will make the decision of either remaining in its current partition set or migrating to a different one. The candidate partition sets for $v$ are those where the highest number of its neighbors are located. Since migrating a vertex potentially introduces a computational overhead, vertex $v$ will preferentially choose to stay in its current partition set if it is one of the candidates. At the end of the iteration, all vertices that decided to migrate will move to their desired partition sets. Furthermore, in order to keep the different partition sets balanced, they have an associated upper capacity. Vertices are allowed to migrate to a different partition set only if its upper capacity has not been yet reached. Vaquero {\em et al.} have shown that  partitioning the vertices of the graph by using the \spinner algorithm speeds up the execution of distributed graph algorithms in the context of Social Network Analysis, Mobile Network Communications, and Bioinformatics~\cite{DBLP:conf/icdcs/VaqueroCLM14}. Based on these experimental findings, we chose to adopt \spinner as partitioning algorithm for \algo.

\subsection{Layout}\label{sse:layout}
The layout step is the core of \algo. To execute it, we need to address the challenges discussed in Section~\ref{sse:challenges}.  We exploit the experimental evidence that in a drawing computed by a force-directed algorithm (see, e.g.,~\cite{Handbook-kob}): 

\begin{itemize}
\item[$(a)$] The graph theoretic distance between two vertices is a good approximation of their geometric distance; 
\item[$(b)$] The repulsive forces between two vertices $u$ and $v$ tend to be less influential as the geometric distance between $u$ and $v$ increases.
\end{itemize} 

\begin{figure}
\centering
\subfigure[]{\includegraphics[width=0.4\columnwidth]{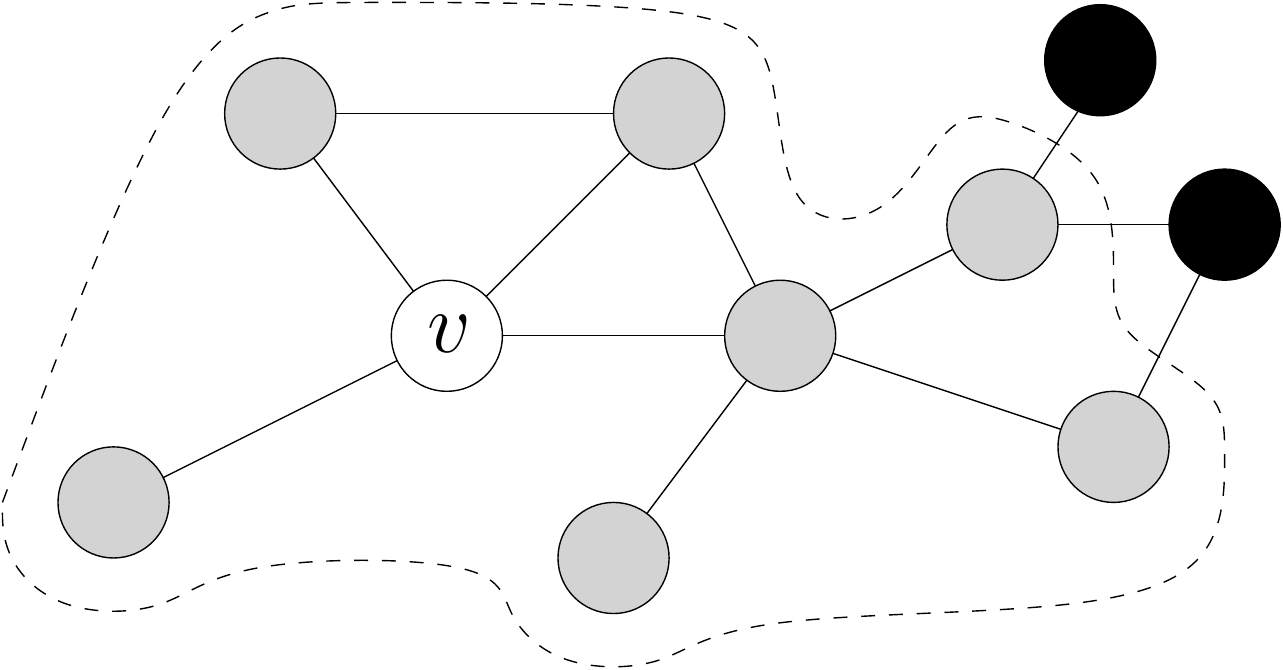}\label{fi:neighborhood}}\hfil
\subfigure[]{\includegraphics[width=0.35\columnwidth]{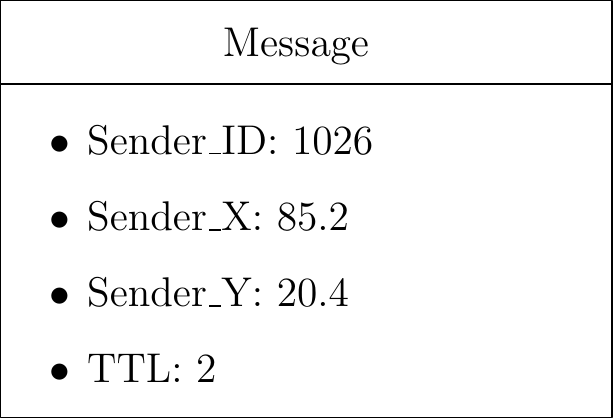}\label{fi:message}}
\caption{(a) The $2$-neighborhood of a vertex $v$ (white). (b) The structure of a message used in the Layout phase.}
\end{figure}

Based on these two observations, we find it reasonable to adopt a locality principle for the repulsive forces. Namely, we assume that, for a suitably defined integer constant $k$, the repulsive force acting on each vertex $v$ only depends on its \emph{$k$-neighborhood} $N_v(k)$, i.e., the set of vertices whose graph theoretic distance from $v$ is at most $k$ (see also Figure~\ref{fi:neighborhood}). Clearly, depending on the structure of the graph, small values of $k$ may affect the accuracy of the forces acting on each vertex. On the other hand, increasing $k$ may cause a very high communication load (see Section~\ref{sse:challenges}); this aspect will be better clarified below. Therefore, finding a good trade-off for the value of $k$ is one of the main goals of our experimental evaluation (Section~\ref{se:experiments}). We also analyze and discuss the impact of $k$ on the theoretical time complexity of the algorithm (Section~\ref{sse:complexity}).  

The attractive and repulsive forces acting on a vertex are defined according to the FR model (see Section~\ref{sse:fr-alg}). In our distributed implementation, each drawing iteration consists of a sequence of Giraph supersteps and works as follows. 

By means of a controlled flooding technique, every vertex $v$ acquires the positions of all vertices in $N_v(k)$. In the first superstep, $v$ sends a message to its neighbors, which contains the coordinates of $v$, its unique identifier, and an integer number, called \emph{TTL (Time-To-Live)}, initially set equal to $k$ (see also Figure~\ref{fi:message}). 

In the second superstep, $v$ processes the received messages and uses them to compute the attractive and repulsive forces with respect to its adjacent vertices. Vertex $v$ uses an efficient data structure $H_v$ (a hash set) to store the unique identifiers of its neighbors. Also, it decreases by one unit the TTL of each received message, and forwards this message to all its neighbors. 

In superstep $i$ ($i > 2$), vertex $v$ processes the received messages and, for each message, $v$ first checks whether the sender $u$ is already present in $H_v$. If not, $v$ uses the message to compute the repulsive force with respect to $u$, and then stores $u$ to $H_v$. Otherwise, the force exerted by $u$ on $v$ had already been computed in some previous superstep. In the first case, $v$ decreases the TTL of the message by one unit and, if it is still greater than zero, the message is broadcasted to its neighbors; otherwise, $v$ discards the message. When no message is sent, the coordinates of each
vertex are updated and the iteration is ended. The amount of messages sent throughout an iteration clearly depends on $k$ and on the sizes of the $k$-neighborhoods of the vertices, which is also related to the diameter of the graph. 

\subsection{Reinsertion}\label{sse:resinsertion}
Once a drawing of the pruned graph has been computed, we reinsert the degree-one vertices by means of an ad-hoc technique. The general idea is as follows. For each vertex $v$ of the pruned graph, we aim at reinserting the degree-one neighbors of $v$ in a geometric neighborhood of $v$, minimizing the interference with other possible edges and vertices.  Consider the circumference $\gamma_v$ centered at $v$ with radius $\rho$, where $\rho$ is some constant fraction of the length of the shortest edge  incident to $v$. We reinsert the degree-one neighbors of $v$ by uniformly distributing them along $\gamma_v$, while avoiding edge overlap. We experimentally tuned $\rho$ to $0.2$.
This reinsertion strategy works fine if the disk determined by $\gamma_v$ does not contain vertices other than $v$. In order to enforce this property as much as possible, the intensity of the repulsive force exerted by $v$ on the other vertices, during the layout step, is proportional to the number of its degree-one neighbors.

\medskip
The pipeline described above is applied independently to each connected component of the input graph.
The layouts of the different components are then conveniently arranged in a matrix, so to avoid overlap.
The pre-processing phase that computes the connected components of the graph is a distributed adaptation of a classical BFS algorithm, based on a simple flooding technique. 

\subsection{Teorethical Time Complexity}\label{sse:complexity}
We conclude the description of our algorithm with the analysis of its asymptotic worst-case time complexity. An experimental evaluation of the practical running time and scalability of \algo is presented in Section~\ref{se:experiments}.

Let $G$ be a graph with $n$ vertices and maximum vertex degree $\Delta$. Recall that $k$ is the integer value used to initialize the TTL of each message. Then the local function computed by each vertex costs $O(\Delta^k)$, since each vertex needs to process (in constant time) one message for each of its neighbors at distance at most $k$, and the number of these neighbors is $O(\Delta^k)$.  Moreover, let $c$ be the number of computing units. Assuming that each of them handles (approximately) $n/c$ vertices, we have that each superstep costs $O(\Delta^k)\frac{n}{c}$. Let $s$ be the maximum number of supersteps that \algo performs (if no equilibrium is reached before), then the time complexity is $O(\Delta^k)s\frac{n}{c}$. If we assume that $c$ and $s$ are constant in the size of the graph, the cost of the algorithm reduces to $O(\Delta^k)n$, which in the worst case corresponds to $O(n^{k+1})$.

\section{Experimental Analysis}\label{se:experiments}

We conducted an experimental evaluation of our algorithm, \algo, in terms of quality of the computed drawings and scalability. We started from the following experimental hypotheses (expectations):

\begin{itemize}

\item[{\tt H1.}] For small values of $k$ ($k \leq 3$), \algo can draw graphs up to one million edges in a reasonable time (few minutes), on a cloud computing platform whose usage cost per hour is relatively low. Also, \algo achieves good performances in terms of weak and strong scalability.  

\item[{\tt H2.}] The quality of the drawings computed by \algo is comparable to that of drawings computed by a Fruchterman-Reingold (FR) centralized algorithm, even for relatively small values of $k$ ($3 \leq k \leq 4$).  

\item[{\tt H3.}] For graphs with a relatively small diameter, small increases of $k$ may give rise to relatively high improvements of the drawing quality. Nevertheless, large graphs with a small diameter may cause a dramatic growth of the running time when $k$ is (even slightly) increased. 

\end{itemize}

Hypotheses {\tt H1} is motivated by the fact that, for small values of $k$, the amount of data stored at each vertex, as well as the message traffic load, should remain limited. Hypothesis {\tt H2} follows from the fact that, for a vertex $v$ and for a relatively small value of $k$, most of the vertices that are not in $N_v(k)$ are far from $v$ in the drawing (i.e., the theoretic distance is a good approximation of the geometric distance). 
About {\tt H3}, we expect that when the diameter of the graph is small, increasing the value of $k$ quickly leads every $N_v(k)$ to include the majority of the vertices of the graph. This should result in a more accurate computation of the repulsive forces but, at the same time, in a significant growth of the traffic load, and hence of the running time, especially when the graph is very large.  

The next subsection discusses some details of our implementation, while Sections~\ref{sse:exp-setting} and~\ref{sse:exp-results} describe the experimental setting and results, respectively.

\subsection{Implementation Details}\label{sse:implementation}

We implemented \algo using the version $1.1$ of the Giraph framework, and the version $2.6$ of the Apache Hadoop framework. The source code is publicly available\footnote{\url{http://www.geeksykings.eu/gila}}.  In what follows, we discuss the implementation and tuning of some salient parameters of the algorithm. These settings are similar to those used in the centralized implementation of the FR algorithm provided by the Open Graph Drawing Framework (OGDF)~\cite{DBLP:reference/crc/ChimaniGJKKM13}, a well-known C++ library already used in several applications and experimental works (see, e.g.,~\cite{DBLP:conf/gd/BartelGKM10,DBLP:conf/apvis/ChimaniJS08,DBLP:conf/apn/HeinerHLRS12,DBLP:journals/tvcg/MuelderM08}).

Initially, the vertices of the input graph are randomly placed within a frame of size $1200 \times 1200$. In the layout step, the forces acting on each vertex are defined according to the FR model (see Section~\ref{sse:fr-alg}): the constant $d$ representing the ideal edge length is defined as $d = N_s+\sqrt{N_h^2+N_w^2}$, where $N_s$, $N_h$, and $N_w$ are three constants all set equal to $20$. They correspond, in the displayed visualization, to the ideal distance between two vertices ($N_s$), to the height ($N_h$) and to the width ($N_w$) of the graphic feature representing a vertex (e.g., a rectangle or a circle). 

Each computation is halted after a superstep if less than $15\%$ of the vertices has a displacement larger than $0.01$ units. This condition is evaluated using an aggregator (see Section~\ref{sse:giraph}). Also, the maximum possible displacement for a vertex is computed independently for each connected component of the graph as follows. Let $\mathcal C$ be a connected component of the input graph $G$. Let $n$ be the number of vertices of $\mathcal C$, and let $a$ be the aspect ratio (i.e., the ratio between the height and the width) of the frame enclosing the initial drawing of $\mathcal C$. The maximum possible vertex displacement at superstep $h$ is set to $d_{\mathcal C}(h) = \sqrt{(\frac{n}{a}d)}{0.93}^h$.

\subsection{Experimental Setting}\label{sse:exp-setting}

In the following we describe: the graph benchmark on which we ran \algo and a centralized spring embedding algorithm, the metrics adopted to evaluate the performance of the algorithms, and the distributed infrastructure used for \algo.

\newcommand{\real}{\texttt{Real}\xspace}
\newcommand{\synther}{\texttt{Synth-Random}\xspace}
\newcommand{\synthba}{\texttt{Synth-ScaleFree}\xspace}

\paragraph{Graph benchmark}
We used three different benchmarks of graphs: 

\begin{itemize}
\item {\real}. It consists of $13$ real networks, with up to $1.5$ million edges. These graphs have been taken from the Sparse Matrix Collection of the University of Florida\footnote{\small\url{http://www.cise.ufl.edu/research/sparse/matrices/}}, from the Stanford Large Networks Dataset Collection \footnote{\small\url{http://snap.stanford.edu/data/index.html}}, and from the Network Data Repository \footnote{\small\url{http://www.networkrepository.com/}}~\cite{nr-aaai15}. Details about name, type, and structure of these graphs are reported in Table~\ref{ta:realgraphs}. Previous experiments on the subject use a comparable number of real graphs (see, e.g.,~\cite{DBLP:conf/egpgv/TikhonovaM08}).

\definecolor{Gray}{gray}{0.9}

\begin{table}[t]
\centering
\renewcommand{\arraystretch}{1.4}
\footnotesize
  
\begin{tabular}{| l | r | r | r | l |}
    \hline
    \rowcolor{Gray}
    \textsc{Name} & $|V|$ & $|E|$ & \textsc{Diameter} & \textsc{Description} \\\hline
	add32 & 4,960 & 9,462 & 28 & circuit simulation problem\\     
    ca-GrQc & 5,242 & 14,496 & 17 & collaboration network\\ 
    grund & 15,575 & 17,427 & 15 & circuit simulation problem \\
    p2p-Gnutella04 & 10,876 & 39,994 & 9 & P2P network\\
    pGp-giantcompo & 10,680 & 48,632 & 17 & email communication network \\
    ca-CondMat & 23,133 & 93,497 & 14 & collaboration network\\
    p2p-Gnutella31 & 62,586 & 147,892 & 11  & P2P network\\ 
    asic-320 & 121,523 & 515,300 & 48 & circuit simulation problem  \\    
    amazon0302 & 262,111 & 899,792 & 32 & co-purchasing network\\
    com-amazon & 334,863 & 925,872 & 44 & co-purchasing network\\
    com-DBLP & 317,080 & 1,049,866 & 21 & collaboration network \\
    roadNet-PA & 1,087,562 & 1,541,514 & 782 & road network \\
    \hline
  \end{tabular}
  %\vspace{-3mm}
  \caption{\small Details for the \real benchmark. Isolated vertices, self-loops, and parallel edges have been removed from the original graphs. The graphs are ordered by increasing number of edges.}\label{ta:realgraphs}
\end{table}

\item {\synther}. It contains $18$ synthetic random graphs generated with the Erd\~{o}s-R\'enyi model~\cite{erdos}. These graphs are divided into six groups of three graphs each, with size  (number of edges)  $m \in \{10^4, 5 \cdot 10^4, 10^5, 10^6, 1.5 \cdot  10^6, 2 \cdot 10^6\}$ and density (number of edges divided by number of vertices) in the range $[2,3]$. 

\item {\synthba}. It contains $18$ synthetic scale-free graphs generated with the Barabasi-Albert model~\cite{barabasi}. Again, these graphs are divided into six groups of three graphs each, with size (number of edges) $m \in \{10^4, 5 \cdot 10^4, 10^5, 10^6, 1.5 \cdot  10^6, 2 \cdot 10^6\}$\footnote{For each sample $m$, the actual number of edges of a graph in this sample is approximately $m$, as the generator does not allow to fix the number of edges in a precise way.} and density in the range $[2,3]$.

\end{itemize}

\paragraph{Metrics}
On each graph of the three benchmarks, we ran \algo with $k \in \{2,3,4\}$ and the (centralized) FR algorithm provided by the Open Graph Drawing Framework (OGDF)~\cite{DBLP:reference/crc/ChimaniGJKKM13}. 
To estimate the resources required by \algo, we measured for each computation both the running time and the cost for using the cloud computing distributed infrastructure. 
To estimate the effectiveness of \algo (i.e., the quality of the computed layouts), we compared its drawings with those computed by the centralized OGDF algorithm, in terms of number of edge crossings, edge length standard deviation, and \emph{similarity} between the ``shape'' of the drawing and the ``structure'' of the input graph (see below).

While number of crossings and edge length standard deviation are frequently used to evaluate the quality of a drawing (see, e.g.,~\cite{DBLP:conf/gd/BartelGKM10,DBLP:journals/jgaa/HachulJ07}), the similarity is a quality metric for large graphs recently introduced by Eades {\em et al.}~\cite{DBLP:conf/gd/EadesHKN15}. The idea behind it is to evaluate the Jaccard sum similarity between the input graph and a proximity graph (in our case the Gabriel Graph) obtained from the set of points representing the vertices in the computed drawing. Briefly, the Jaccard sum similarity measures for each vertex $v$ the number of edges incident to $v$ that are shared by the input graph and by the proximity graph (see~\cite{DBLP:conf/gd/EadesHKN15} for more details). For the sake of presentation, and in order to compare the different values, we normalized the data between 0 and 1 for each graph.

\paragraph{Distributed infrastructure}
The experiments on \algo were executed on the Amazon EC2 infrastructure. We experimented three clusters of machines, consisting of $10$, $15$, and $20$ machines, respectively. Each machine is a memory-optimized instance (R3.xlarge) with 4 vCPUs and $30.5$ GiB RAM. The cost per hour to use this infrastructure is about $0.5$ USD per machine. 
%The experiments on the FR algorithm were executed on an Intel i7 3630QM laptop, with 2.4 GHz and 8GB of RAM (the running times obtained for this centralized algorithm are not interesting for our experimental analysis).   

\subsection{Experimental Results}\label{sse:exp-results}

In the following we present and discuss the results of our experiments. In each table used to report the experimental data, the symbol $*$ indicates that the computation on the corresponding instance was halted, either because it took more than 5 hours or because it caused out-of-memory errors.

\paragraph{Running time and cost} 
Tables~\ref{ta:real-results-time}--~\ref{ta:synther-results-time} report the running time and the infrastructure cost of \algo for the the three graph benchmarks. For every instance we executed the algorithm on each of the three clusters of machines, and for each $k=2,3,4$. All computations for the same instance started with the same initial (random) configuration for the vertices.

\begin{figure}
\centering
\subfigure[]{\includegraphics[width=0.49\columnwidth]{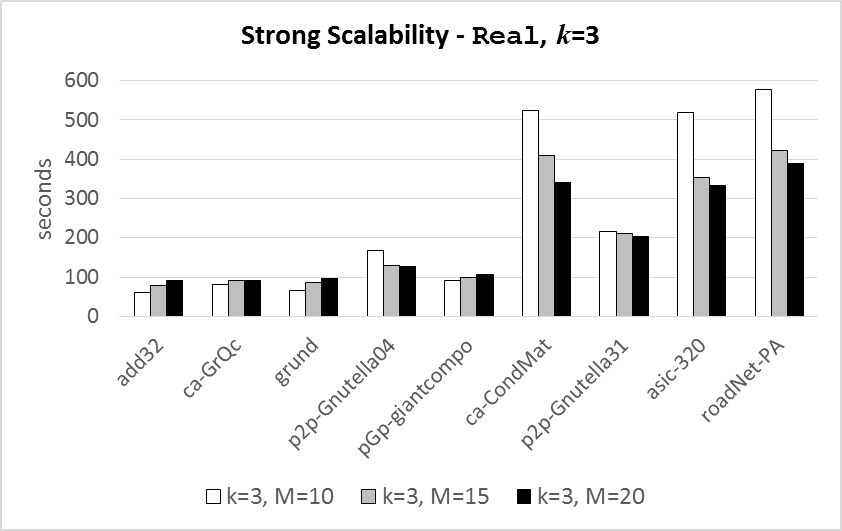}\label{fi:chart-real-ss-k3}}
\subfigure[]{\includegraphics[width=0.49\columnwidth]{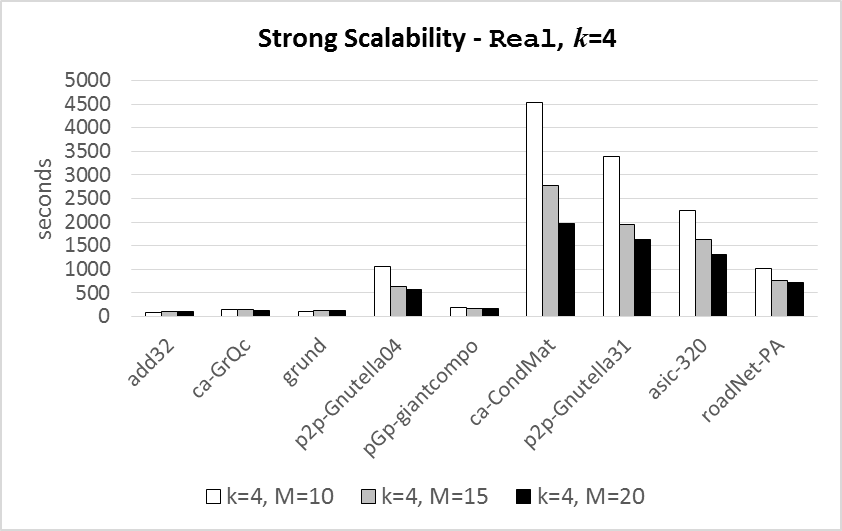}\label{fi:chart-real-ss-k4}}
\caption{Strong scalability on the \real instances for $k=3$ and $k=4$. $M$ denotes the number of machines. The chart reports only the instances that have been successfully computed for both values of $k$.}\label{fi:chart-real-ss}
\end{figure}

Concerning the {\real} graphs (Table~\ref{ta:real-results-time}), it can be seen that for $k=2$ all instances were successfully computed on the smallest cluster of machines, i.e., using just $10$ machines. The computations on the instances with less than $1$ million edges required from $48$ seconds to $4$ minutes, and two of the graphs with more than $1$ million edges were processed in less than $8$ minutes.   
For $k=3$ and $k=4$, the computations became harder, although the percentages of instances successfully computed remained high, and increased using more machines. In general, \algo exhibits a good strong scalability. We recall that the strong scalability of a distributed algorithm is a measure of how much the algorithm improves its time performance when we increase the number of machines on a given instance. The time improvement is mostly evident for $k=3$ and $k=4$, and on the largest instances of the benchmark (see also Figure~\ref{fi:chart-real-ss}). For most of these instances, the time reduction passing from 10 machines to 20 machines was more than $30\%$; for the graph amazon0302 (not reported in Figure~\ref{fi:chart-real-ss}), the time reduction was greater than $50\%$ with $k=3$. For the smaller instances, using the small cluster of machines was often conveniently, since the fixed computational (time and space) resources required by the infrastructure were not adequately amortized over the size of the graph.

\definecolor{Gray}{gray}{0.9}
\definecolor{DGray}{gray}{0.8}

\begin{table}
\centering
\renewcommand{\arraystretch}{1.3}
\scriptsize
\begin{tabular}{| l | l | r | r | r | r | r | r | }
    \hline
    \rowcolor{DGray}
     & & \multicolumn{2}{c|}{\algo - $k=2$} & \multicolumn{2}{c|}{\algo - $k=3$} & \multicolumn{2}{c|}{\algo - $k=4$}\\
    \cline{2-8} 
    \rowcolor{Gray}   
    & \textsc{Graph Name} & \textsc{Time [sec.]} & \$ &  \textsc{Time [sec.]} & \$  &  \textsc{Time [sec.]} & \$ \\\hline % &  Time [sec.] & \$  \\\hline
\multirow{13}{*}{\rotatebox[origin=c]{90}{10 machines}} & add32 & 48 & 0.06 & 61 & 0.08 & 81 & 0.10\\% & 144 & 0.14 \\
& ca-GrQc & 50 & 0.06 & 81 & 0.10 & 154 & 0.19\\% & 263 & 0.35 \\
& grund & 49 & 0.06 & 65 & 0.08 & 101 & 0.12\\% & 563 & 0.19 \\
& p2p-Gnutella04 & 56 & 0.07 & 167 & 0.21 & 1051 & 1.30\\% & 225 & 4.58 \\
& pGp-giantcompo & 52 & 0.06 & 90 & 0.11 & 185 & 0.23\\% & 210 & 0.52 \\
& ca-CondMat & 81 & 0.10 & 523 & 0.65 & 4526 & 5.58\\% & $*$ & $*$ \\
& p2p-Gnutella31 & 75 & 0.09 & 217 & 0.27 & 3379 & 4.17\\% & $*$ & $*$ \\ 
& asic-320 & 132 & 0.31 & 519 & 0.64 & 2245 & 2.77\\% & $*$ & $*$ \\
& amazon0302 & 235 & 0.09 & 1703 & 2.10 & $*$ & $*$\\%  & $*$ & $*$ \\
& com-Amazon & 251 & 0.29 & 1079 & 1.33 & $*$ & $*$\\% & $*$ & $*$ \\
& com-DBLP & 468 & 0.16 & $*$ & $*$ & $*$ & $*$\\% & $*$ & $*$ \\
& roadNet-PA & 347 & 0.43 & 576 & 0.71 & 1011 & 1.25\\% & $*$ & $*$ \\ 
    \hline
    \hline
    \multirow{13}{*}{\rotatebox[origin=c]{90}{15 machines}} & add32 & 60 & 0.11 & 79 & 0.14 & 99 & 0.18\\% & 129 & 0.23 \\
& ca-GrQc & 61 & 0.11 & 91 & 0.17 & 142 & 0.26\\% & 225 & 0.41 \\
& grund & 63 & 0.11 & 85 & 0.15 & 117 & 0.21\\% & 163 & 0.30 \\
& p2p-Gnutella04 & 67 & 0.12 & 130 & 0.24 & 643 & 1.17\\% & 1972 & 3.59 \\
& pGp-giantcompo & 66 & 0.12 & 100 & 0.18 & 173 & 0.32\\% & 330 & 0.60 \\
& ca-CondMat & 87 & 0.16 & 408 & 0.74 & 2765 & 5.03\\% & $*$ & $*$ \\
& p2p-Gnutella31 & 74 & 0.13 & 211 & 0.38 & 1946 & 3.54\\% & $*$ & $*$ \\ 
& asic-320 & 110 & 0.20 & 352 & 0.64 & 1622 & 2.95\\% & $*$ & $*$ \\
& amazon0302 & 184 & 0.34 & 1011 & 1.84 & $*$ & $*$\\% & $*$ & $*$ \\
& com-amazon & 203 & 0.37 & 829 & 1.51 & 3188 & 5.80\\% & $*$ & $*$ \\
& com-DBLP & 375 & 0.68 & $*$ & $*$ & $*$ & $*$\\% & $*$ & $*$ \\
& roadNet-PA & 276 & 0.50 & 421 & 0.77 & 766 & 1.39\\% & $*$ & $*$ \\
    \hline
        \hline
\multirow{13}{*}{\rotatebox[origin=c]{90}{20 machines}} & add32 & 64 & 0.15 & 90 & 0.22 & 102 & 0.25\\% & 129 & 0.23 \\
& ca-GrQc & 63 & 0.15 & 90 & 0.22 & 130 & 0.31\\% & 225 & 0.41 \\
& grund & 65 & 0.16 & 96 & 0.23 & 119 & 0.29\\% & 163 & 0.30 \\
& p2p-Gnutella04 & 72 & 0.17 & 128 & 0.31 & 575 & 1.38\\% & 1972 & 3.59 \\
& pGp-giantcompo & 71 & 0.17 & 106 & 0.26 & 173 & 0.42\\% & 330 & 0.60 \\
& ca-CondMat & 88 & 0.21 & 341 & 0.82 & 1967 & 4.74\\% & $*$ & $*$ \\
& p2p-Gnutella31 & 94 & 0.23 & 203 & 0.49 & 1623 & 3.91\\% & $*$ & $*$ \\ 
& asic-320 & 131 & 0.32 & 333 & 0.80 & 1306 & 3.15\\% & $*$ & $*$ \\
& amazon0302 & 172 & 0.41 & 842 & 2.03 & $*$ & $*$\\% & $*$ & $*$ \\
& com-amazon & 177 & 0.43 & 601 & 1.45 & 2653 & 6.39\\% & $*$ & $*$ \\
& com-DBLP & 364 & 0.88 & 4,711 & 11.35 & $*$ & $*$\\% & $*$ & $*$ \\
& roadNet-PA & 264 & 0.64 & 390 & 0.94 & 713 & 1.72\\% & $*$ & $*$ \\
    \hline
  \end{tabular}
  \caption{\small Running time and infrastructure cost for the \real benchmark, on the three types of  clusters.}\label{ta:real-results-time}
\end{table}

\begin{table}
\centering
\renewcommand{\arraystretch}{1.3}
\footnotesize
\begin{tabular}{| l | r | r | r | r | r | r | r | }
    \hline
    \rowcolor{DGray}
     & & \multicolumn{2}{c|}{\algo - $k=2$} & \multicolumn{2}{c|}{\algo - $k=3$} & \multicolumn{2}{c|}{\algo - $k=4$}\\ \cline{2-8}
    \rowcolor{Gray}
    & $|E|$ & \textsc{Time [sec.]} & \$ &  \textsc{Time [sec.]} & \$  &  \textsc{Time [sec.]} & \$ \\\hline % &  Time [sec.] & \$  \\\hline
\multirow{6}{*}{\rotatebox[origin=c]{90}{10 machines}} & 10,000 & 44 & 0.05 & 59 & 0.07 & 91 & 0.11 \\
& 50,000 & 52 & 0.06 & 79 & 0.10 & 162 & 0.20 \\
& 100,000 & 60 & 0.07 & 111 & 0.14 & 381 & 0.47\\
& 1,000,000 & 205 & 0.25 & 751 & 0.93 & $*$ & $*$ \\
& 1,500,000 & 302 & 0.37 & 1,293 & 1.59 & $*$ & $*$ \\
& 2,000,000 & 452 & 0.56 & 2,088 & 2.58 & $*$ & $*$ \\
 
    \hline
    \hline
    \multirow{6}{*}{\rotatebox[origin=c]{90}{15 machines}} & 10,000& 54 & 0.10 & 70 & 0.13 & 97 & 0.18 \\
& 50,000 & 60 & 0.11 & 86 & 0.16 & 147 & 0.27 \\
& 100,000 & 73 & 0.13 & 120 & 0.22 & 319 & 0.58 \\
& 1,000,000 & 171 & 0.31 & 540 & 0.98 & 2,658 & 4.84 \\
& 1,500,000 & 235 & 0.43 & 995 & 1.81 & $*$ & $*$ \\
& 2,000,000 & 281 & 0.51 & 1,504 & 2.74 & $*$ & $*$ \\

    \hline
        \hline
\multirow{6}{*}{\rotatebox[origin=c]{90}{20 machines}} & 10,000& 62 & 0.15 & 80 & 0.19 & 107 & 0.26 \\
& 50,000 &77 & 0.18 & 106 & 0.26 & 161 & 0.39 \\
& 100,000 &90 & 0.22 & 134 & 0.32 & 286 & 0.69 \\
& 1,000,000 &167 & 0.40 & 447 & 1.08 & 2,113 & 5.03 \\
& 1,500,000 &209 & 0.50 & 809 & 1.95 & 4,303 & 10.36 \\
& 2,000,000 &271 & 0.65 & 1,154 & 2.78 & $*$ & $*$ \\

    \hline
  \end{tabular}
  \caption{\small Running time and infrastructure cost for the \synther benchmark achieved on the three clusters. For each value of $|E|$ we reported the average value on the three graphs with $|E|$ edges.}\label{ta:synther-results-time}
\end{table}

\begin{table}
\centering
\renewcommand{\arraystretch}{1.3}
\footnotesize
\begin{tabular}{| l | r | r | r | r | r | r | r | }
    \hline
    \rowcolor{DGray}
     & & \multicolumn{2}{c|}{\algo - $k=2$} & \multicolumn{2}{c|}{\algo - $k=3$} & \multicolumn{2}{c|}{\algo - $k=4$}\\ \cline{2-8}
     \rowcolor{Gray}
    & $|E|$ & \textsc{Time [sec.]} & \$ &  \textsc{Time [sec.]} & \$  &  \textsc{Time [sec.]} & \$ \\\hline % &  Time [sec.] & \$  \\\hline
\multirow{6}{*}{\rotatebox[origin=c]{90}{10 machines}} & 10,000 & 47 & 0.06 & 78 & 0.10 & 215 & 0.26 \\
& 50,000 &58 & 0.07 & 227 & 0.28 & 1,857 & 2.29 \\
& 100,000 &77 & 0.10 & 603 & 0.74 & 4,035 & 4.98 \\
& 1,000,000 &743 & 0.92 & $*$ & $*$ & $*$ & $*$ \\
& 1,500,000 &1,055 & 1.30 & $*$ & $*$ & $*$ & $*$ \\
& 2,000,000 &1,689 & 2.08 & $*$ & $*$ & $*$ & $*$ \\
 
    \hline
    \hline
    \multirow{6}{*}{\rotatebox[origin=c]{90}{15 machines}} & 10,000 & 54 & 0.10 & 84 & 0.15 & 173 & 0.32 \\
&  50,000 &70 & 0.13 & 185 & 0.34 & 1,174 & 2.14 \\
&  100,000 &87 & 0.16 & 398 & 0.73 & 2,747 & 5.00 \\
&  1,000,000 &476 & 0.87 & $*$ & $*$ & $*$ & $*$ \\
&  1,500,000 &766 & 1.39 & $*$ & $*$ & $*$ & $*$ \\
&  2,000,000 &1,164.00 & 2.12 & $*$ & $*$ & $*$ & $*$ \\

    \hline
        \hline
\multirow{6}{*}{\rotatebox[origin=c]{90}{20 machines}} & 10,000 & 63 & 0.15 & 95 & 0.23 & 172 & 0.41 \\
&  50,000 &76 & 0.18 & 180 & 0.43 & 915 & 2.20 \\
&  100,000 &91 & 0.22 & 355 & 0.86 & 2,112 & 5.09 \\
&  1,000,000 &403 & 0.97 & $*$ & $*$ & $*$ & $*$ \\
&  1,500,000 &571 & 1.38 & $*$ & $*$ & $*$ & $*$ \\
&  2,000,000 &956 & 2.31 & $*$ & $*$ & $*$ & $*$ \\

    \hline
  \end{tabular}
  \caption{\small Running time and infrastructure cost for the \synthba benchmark achieved on the three clusters. For each value of $|E|$ we reported the average value on the three graphs with $|E|$ edges.}\label{ta:synthba-results-time}
\end{table}

The data for the \synther and the \synthba graphs are reported in Table~\ref{ta:synther-results-time} and Table~\ref{ta:synthba-results-time}, respectively. From the structural point of view, the    
\synther have a more uniform vertex-degree distribution, while the vertex-degree distribution of the \synthba instances follows a power-law, as they are scale-free graphs. Clearly, vertices of very high degree may cause a significant work load for {\algo}, especially if the graph has a small diameter. Indeed, one can see that \synthba graphs are the most difficult instances for {\algo}: for $k \geq 3$, the algorithm failed the computations on graphs with more than one million edges. Conversely, \algo successfully computed all the \synther graphs for $k \leq 3$ and some of the instances for $k=4$. 

\begin{figure}
\centering
\subfigure[]{\includegraphics[width=0.49\columnwidth]{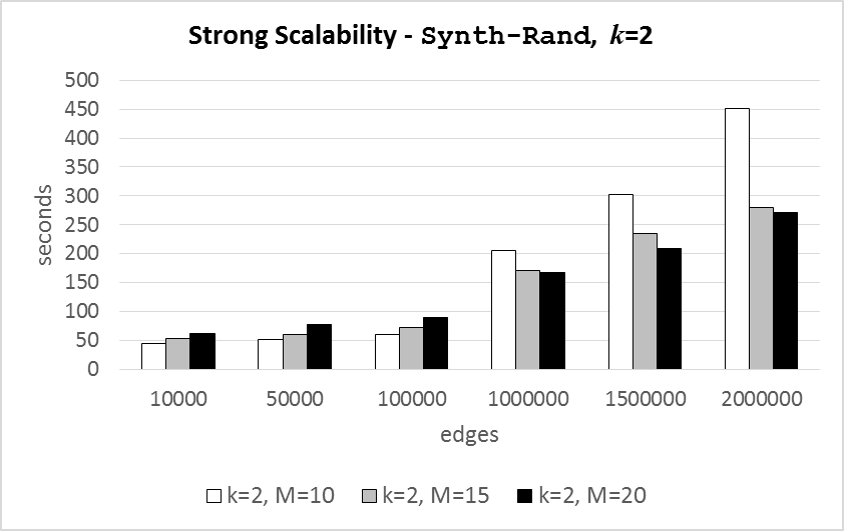}\label{fi:chart-synth-rand-ss-k2}}
\subfigure[]{\includegraphics[width=0.49\columnwidth]{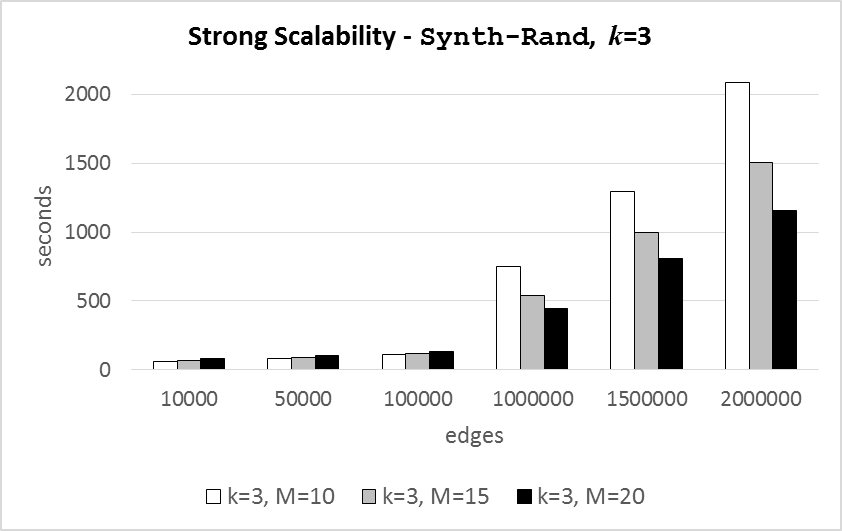}\label{fi:chart-synth-rand-ss-k3}}
\subfigure[]{\includegraphics[width=0.49\columnwidth]{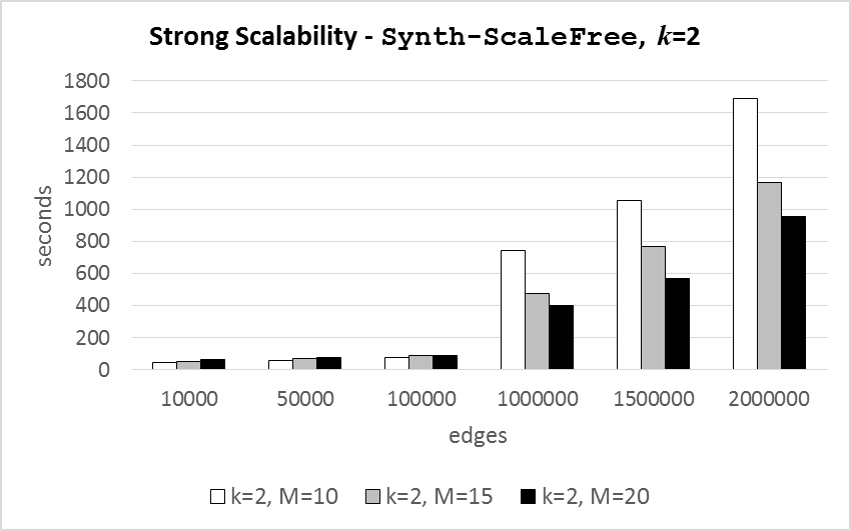}\label{fi:chart-synth-scalefree-ss-k2}}
\subfigure[]{\includegraphics[width=0.49\columnwidth]{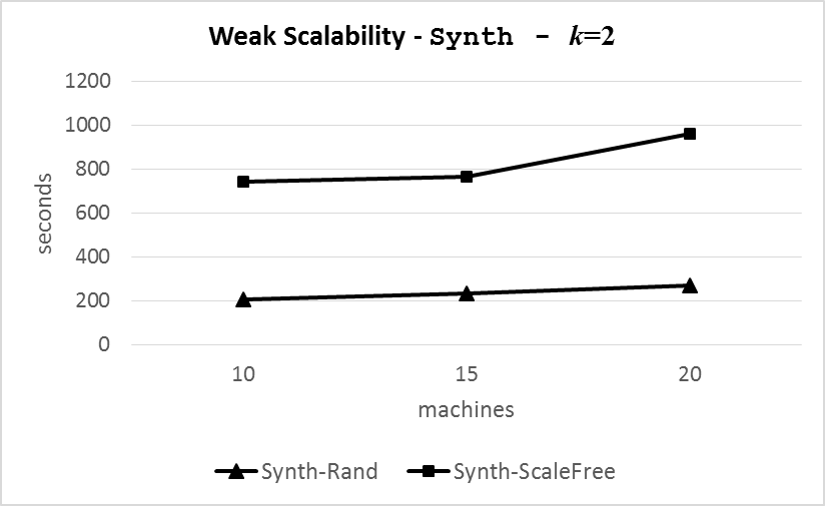}\label{fi:chart-synth-ws-k2}}

\caption{Strong and weak scalability on the \synther and \synthba instances. $M$ denotes the number of machines.}\label{fi:chart-synth-sc-ws}
\end{figure}

The data about strong scalability on the synthetic graphs confirmed the behavior on the \real graphs. Figures~\ref{fi:chart-synth-rand-ss-k2},~\ref{fi:chart-synth-rand-ss-k3}, and~\ref{fi:chart-synth-scalefree-ss-k2} summarize these data for $k=2,3$ on the \synther graphs and for $k=2$ on the \synthba graphs. Again, augmenting the number of machines does not help when the graphs are relatively small, while it dramatically reduces the running time on the largest instances (of about $40\%$ in the average).
We also report a chart about the weak scalability of {\algo} on the synthetic graphs (see Figure~\ref{fi:chart-synth-ws-k2}). We recall that the weak scalability of a distributed algorithm estimates how the running time varies when we increase the number of machines and the size of the instances, while keeping the portion of instance handled by each machine constant. To this aim, we examined the behavior of \algo for the instances with $10^6$ edges on $10$ machines, with $1.5 \cdot 10^6$ edges on $15$ machines, and with $2 \cdot 10^6$ edges on $20$ machines. Thus, the number of edges per machine remained approximately equal to $100,000$ (since the graphs have a similar density, we can also assume that the number of vertices per machine remained approximately the same). In the chart of Figure~\ref{fi:chart-synth-ws-k2} we summarize the results for $k=2$, for which we have a complete set of data. For the \synther graphs, the time increased only by $14-15\%$ passing from a sample to the next (a constant time value would be the optimum). For the \synthba graphs the time was still rather stable passing from $10$ machines ($1,000,000$ vertices) to $15$ machines ($1,500,000$ vertices), but it increased of about $25\%$ passing from $15$ to $20$ machines ($2,000,000$ nodes). This suggests that the weak scalability on scale-free graphs is more difficult to predict.

Overall, the results about running time and infrastructure cost seem to largely confirm \texttt{H1}.  

\paragraph{Quality metrics} Table~\ref{ta:real-results-cr} reports the quality metrics of \algo with $k=\{2,3, 4\}$ and of the centralized FR algorithm available in the OGDF library (OGDF-FR in the table). The total number of crossings is divided by the number of edges of the graph, thus indicating the average number of crossings per edge (CRE). Also, for each graph, the series of similarity values (SIM) obtained with the different algorithms and settings has been normalized and scaled between $0$ and $1$, so that $1$ corresponds to the best value in the series. We remark that a drawing of the last five largest graphs in this benchmark could not be computed neither by \algo when $k=4$ nor by OGDF-FR; hence we do not report the corresponding rows in the table. 

\begin{table}
\centering
\renewcommand{\arraystretch}{1.3}
\footnotesize
\begin{tabular}{| l | r | r | r | r | r | r | }
    \hline
    \rowcolor{DGray}
     & \multicolumn{3}{c|}{\algo - $k=2$} & \multicolumn{3}{c|}{\algo - $k=3$} \\ \cline{2-7}
	\rowcolor{Gray}    
    \textsc{Graph Name} & CRE & ELD & SIM & CRE & ESD & SIM \\\hline
add32 & 0.07 & 29.84 & 0.87 & 0.07 & 31.65 & 0.84 \\
ca-GrQc & 1.02 & 15.04 & 0.71 & 0.81 & 25.93 & 0.84 \\
grund & 0.25 & 26.75 & 0.62 & 0.13 & 31.99 & 0.83 \\
p2p-Gnutella04 & 70.46 & 15.72 & 0.26 & 55.17 & 35.03 & 0.65 \\
pgp-giantcompo & 1.62 & 29.28 & 0.70 & 1.27 & 36.88 & 0.87 \\
ca-CondMat & 122.73 & 18.39 & 0.39 & 80.84 & 42.53& 0.51 \\
p2p-Gnutella31 & 654.09 & 16.97 & 0.08 & 544.16 & 36.57 & 0.39 \\
asic-320 & 74.39 & 94.45 & 0.91 & 75.98 & 94.06 & 0.90 \\
amazon0302 & 2,179.66 & 52.28 & 1.00 & 2,136.00 & 54.76 & 0.99 \\
com-amazon & 1,012.53 & 166.06 & 1.00 & 1,005.40 & 165.18 & 0.99 \\
com-DBLP & 10,630.69 & 49.49 & 0.94 & 7,720.00 & 63.99 & 1.00 \\
roadNet-PA  & 911.62 & 370.86 & 1.00 & 913.07 & 370.95 & 0.99 \\
    \hline
    \rowcolor{DGray}
     & \multicolumn{3}{c|}{\algo - $k=4$} & \multicolumn{3}{c|}{OGDF-FR} \\ \cline{2-7}
	\rowcolor{Gray}    
     \textsc{Graph Name} & CRE & ELD & SIM & CRE & ELD & SIM  \\\hline
add32 & 0.05 & 35.84 & 1.00 & 0.09 & 105.62 & 0.44 \\
ca-GrQc & 0.68 & 41.12 & 1.00 & 0.84 & 80.89 & 0.78 \\
grund & 0.10 & 44.24 & 1.00 & 0.33 & 182.17 & 0.04 \\
p2p-gnutella04 & 52.68 & 59.71 & 1.00 & 62.15 & 79.27 & 0.03 \\
pGp-giantcompo & 1.13 & 49.52 & 1.00 & 1.87 & 151.42 & 0.23 \\
ca-CondMat & 64.98 & 73.37 & 0.83 & 78.91 & 118.74 & 1.00 \\
p2p-Gnutella31 & 427.82 & 73.80 & 1.00 & 601.98 & 148.11 & 0.04 \\
asic-320 & 61.60 & 98.44 & 1.00 & $*$ & $*$ & $*$ \\
%amazon0302 & $*$ & $*$ & $*$ & $*$ & $*$ & $*$ & $*$ & $*$ \\
com-amazon & 1,072.55 & 155.64 & 0.97 & $*$ & $*$ & $*$ \\
%com-DBLP & $*$ & $*$ & $*$ & $*$ & $*$ & $*$ & $*$ & $*$ \\
roadNet-PA  & 910.39 & 369.54 & 0.99 & $*$ & $*$ & $*$ \\
    \hline
  \end{tabular}
  \caption{\small Average number of crossings per edge (CRE), edge length standard deviation (ELD), and similarity (SIM) for the \real benchmark. For each graph, the similarity values have been normalized and scaled between $0$ and $1$, so that $1$ corresponds to the best value.}\label{ta:real-results-cr}
\end{table}

\begin{figure}
\centering
\subfigure[]{\includegraphics[width=0.5\columnwidth]{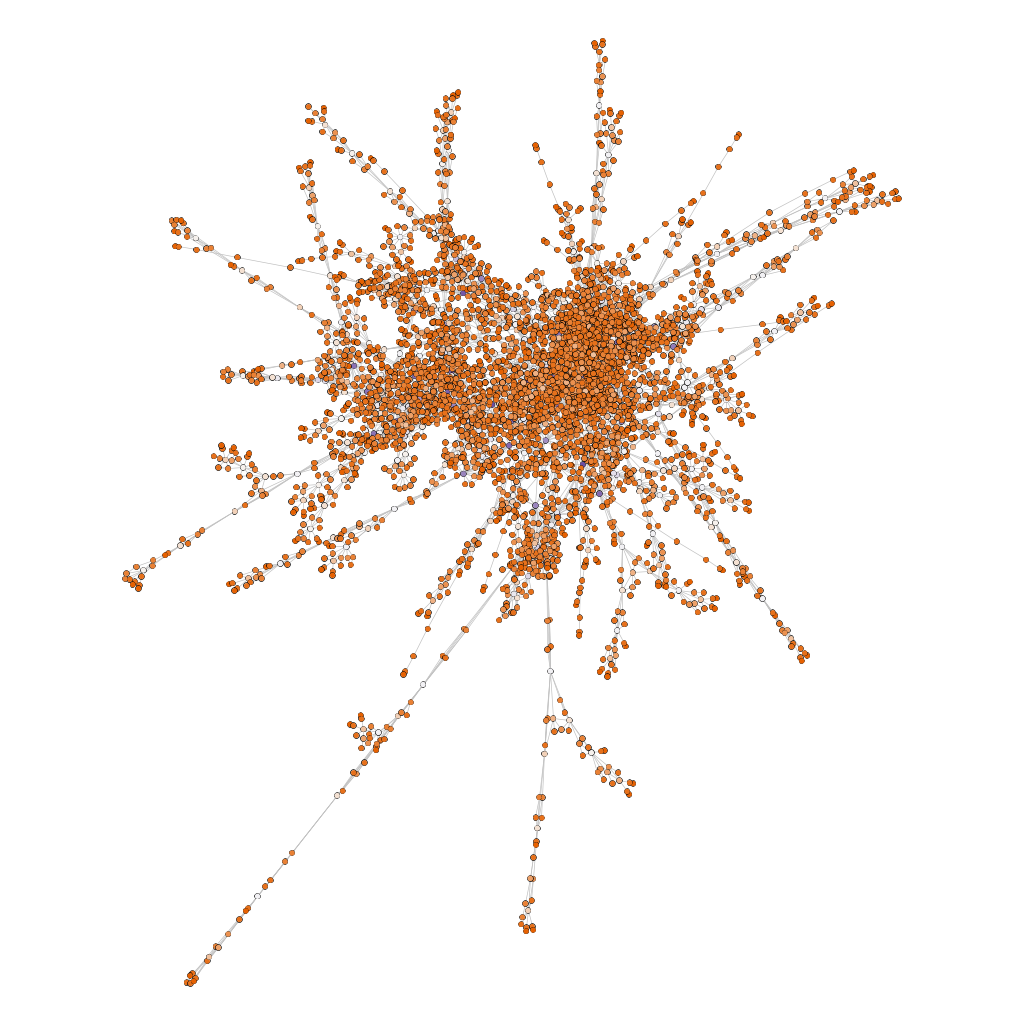}\label{fi:add32_K2}}\hfil
\subfigure[]{\includegraphics[width=0.5\columnwidth]{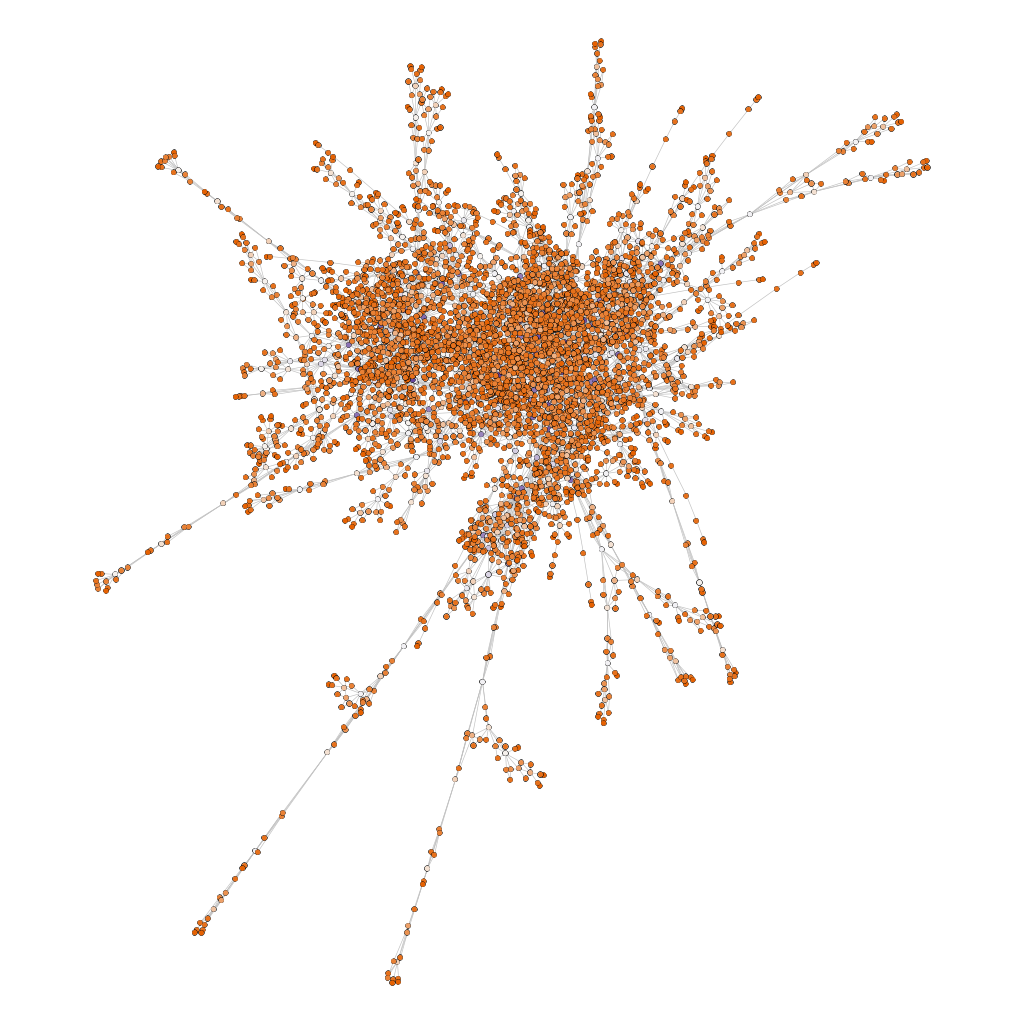}\label{fi:add32_K3}}
\subfigure[]{\includegraphics[width=0.5\columnwidth]{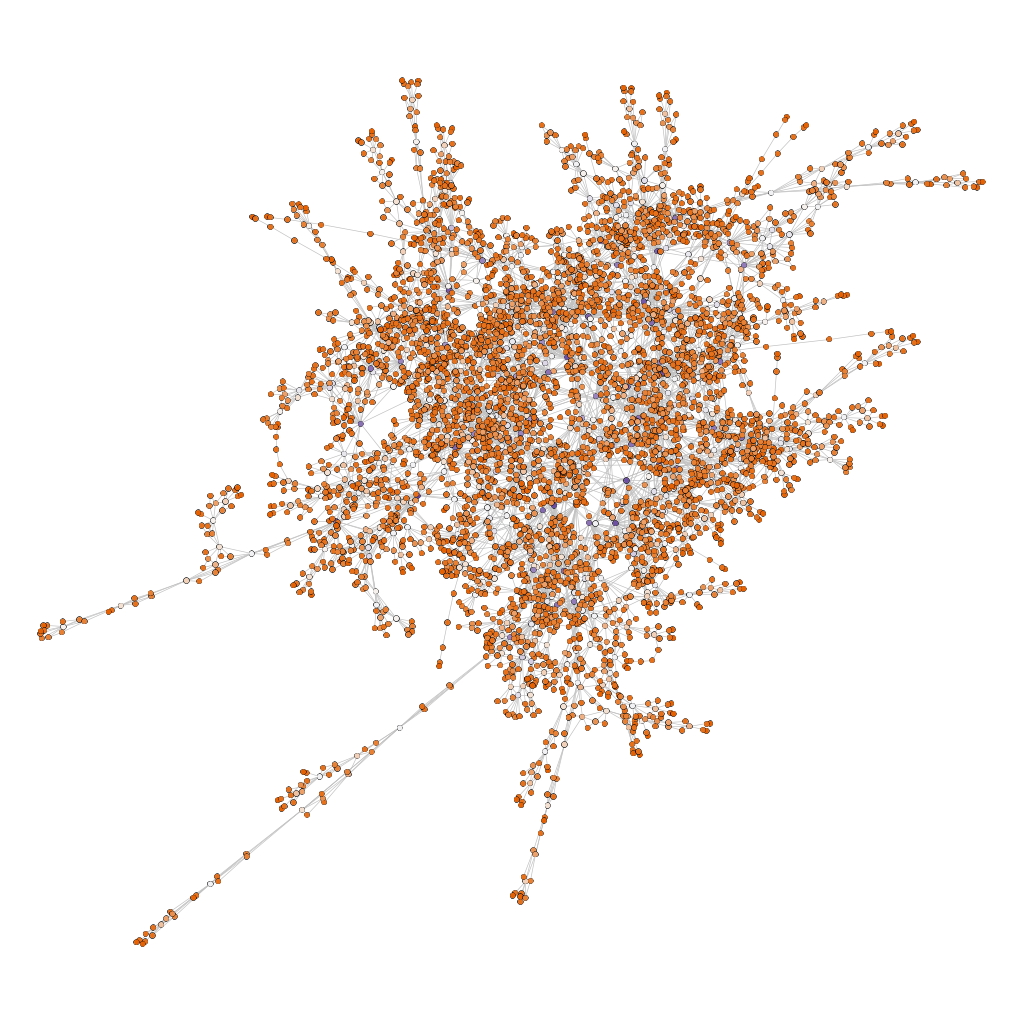}\label{fi:add32_K4}}\hfil
\subfigure[]{\includegraphics[width=0.5\columnwidth]{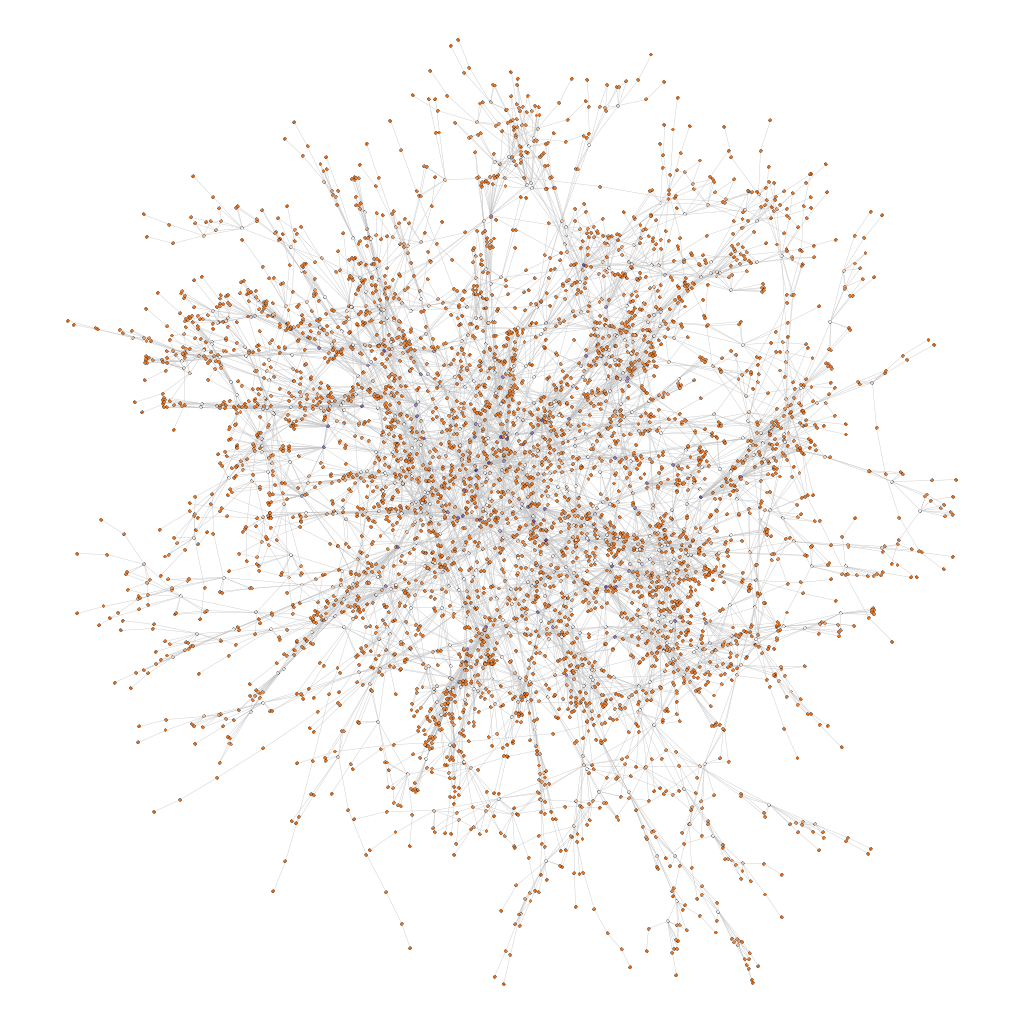}\label{fi:add32_FR}}
\caption{Drawings of the graph add32 computed by \algo, with: (a) $k=2$, (b) $k=3$, and (c) $k=4$. The drawing in (d) has been computed by OGDF-FR.}\label{fi:add32}
\end{figure}

\begin{figure}
\centering
\subfigure[grund - \algo]{\includegraphics[width=0.5\columnwidth]{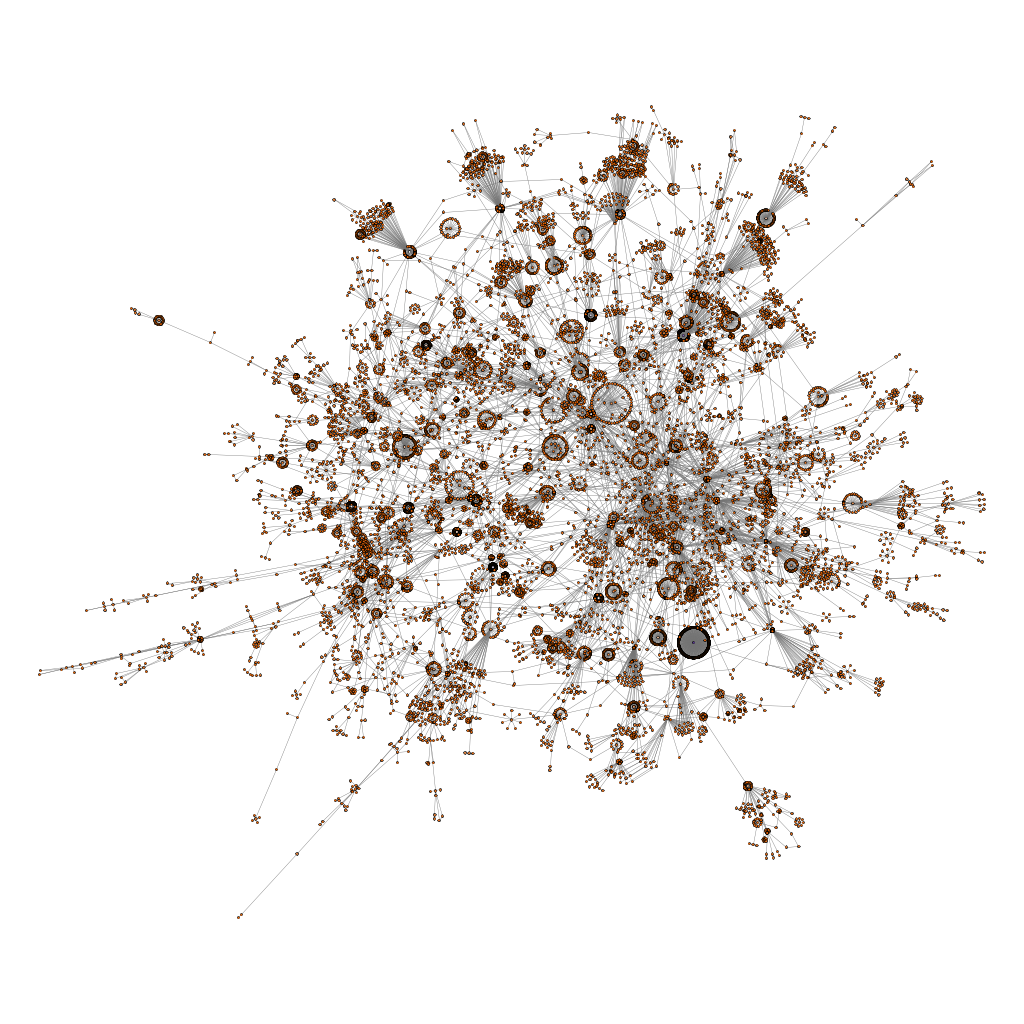}\label{fi:Grund_K5}}\hfil
\subfigure[grund - OGDF-FR]{\includegraphics[width=0.5\columnwidth]{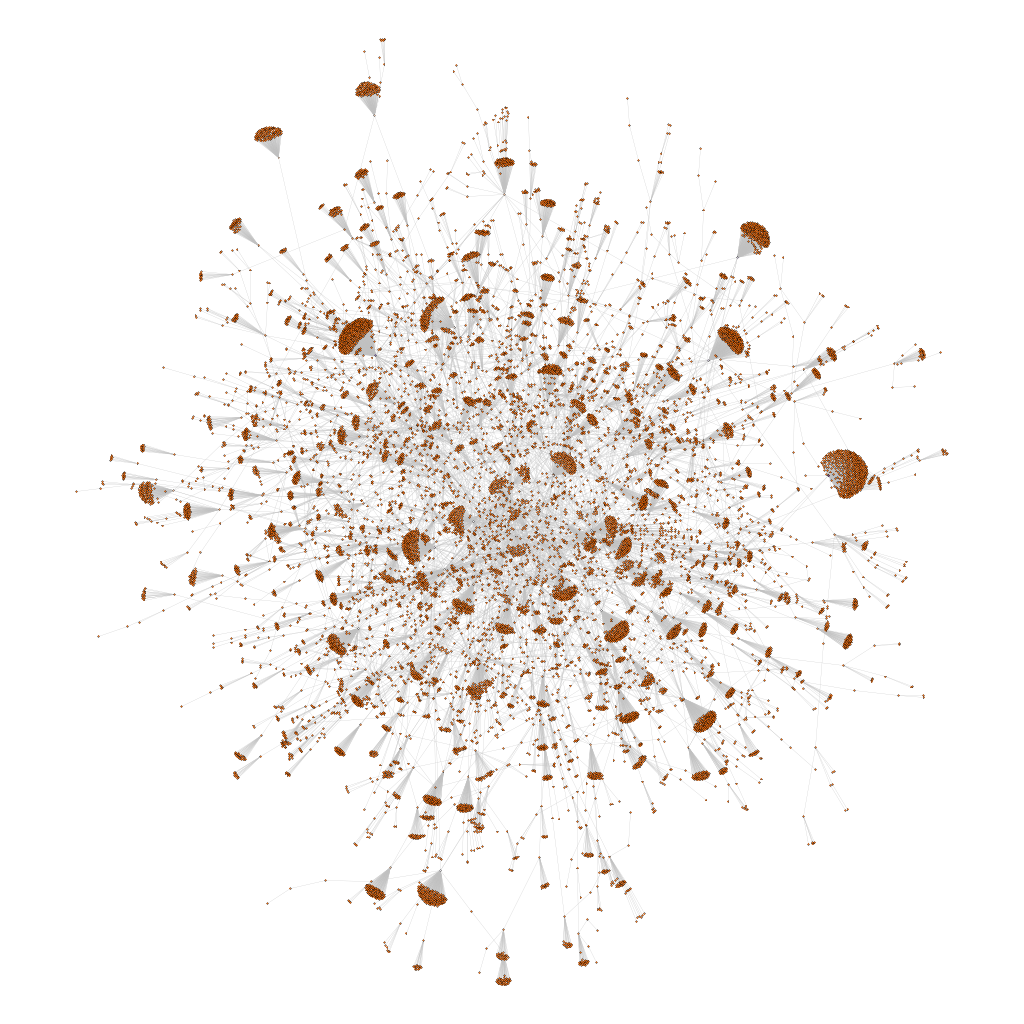}\label{fi:Grund_FR}}
\subfigure[pGp-giantcompo - \algo]{\includegraphics[width=0.5\columnwidth]{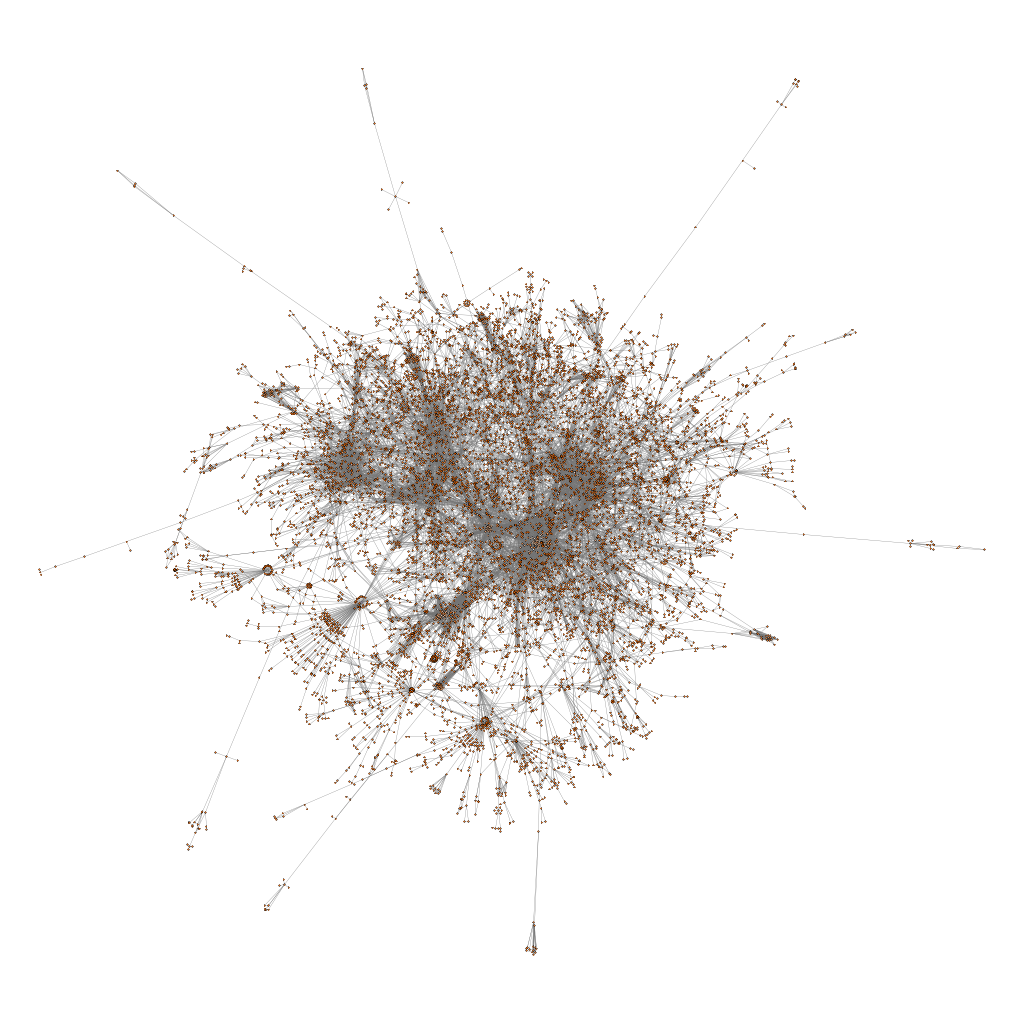}\label{fi:PGPGiantCompo_K5}}\hfil
\subfigure[pGp-giantcompo - OGDF-FR]{\includegraphics[width=0.5\columnwidth]{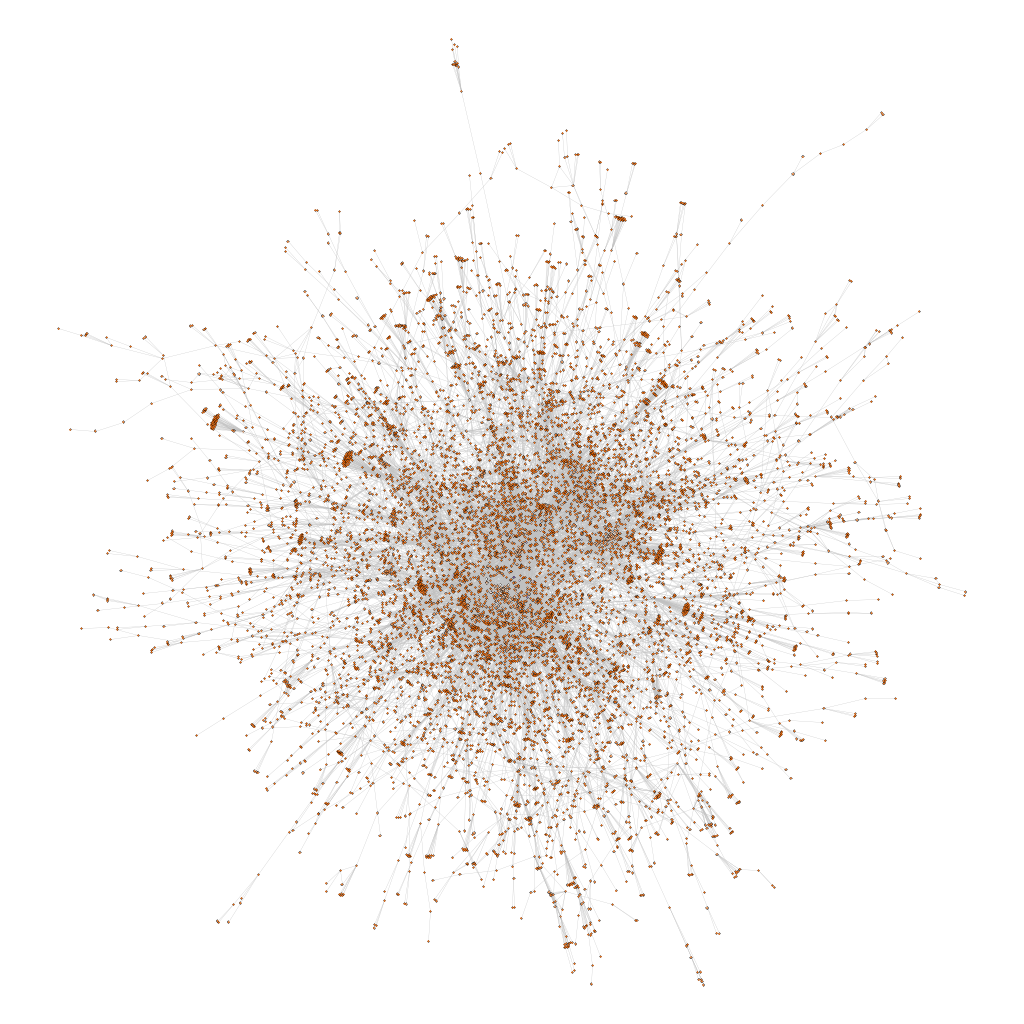}\label{fi:PGPGiantCompo_FR}}
\caption{Drawings of the graphs grund and pGp-giantcompo computed by \algo with $k=5$ and by OGDF-FR.}\label{fi:PGPGiantCompo}
\end{figure}

We first observe that, for almost all instances, the quality of the drawing in terms of crossings and similarity improves when $k$ increases. Recall that for higher values of $k$ the computation of the repulsive forces becomes more precise. For all graphs with less than one million edges, the number of crossings per edge is reduced on average by $34.5\%$, while passing from $k=2$ to $k=4$; also, for $k=4$ we get the best similarity value in all cases. For the largest graphs these two measures are more stable, although we still observe a crossing reduction of about $27.4\%$ on the graph com-DBLP, while passing from $k=2$ to $k=3$. 

Furthermore, on all instances, \algo with $k=4$ behaved better than OGDF-FR in terms of number of crossings; the average improvement is about $33.5\%$. With $k=3$, \algo still produced drawings with less crossings than OGDF-FR in all graphs except one, with average improvement of about $23.2\%$. With $k=2$, \algo caused less crossings than OGDF-FR in the $43\%$ of instances ($3$ over $7$). Concerning the similarity, \algo achieved better values than OGDF-FR in most of the cases, even with $k=2$. 

About the edge length standard deviation, the values of the drawings computed by \algo are always smaller than those of the drawings computed by OGDF-FR. However, these values tend to grow when $k$ increases (except for two of the largest graphs, com-amazon and roadNet-PA). Indeed, for $k=2$, the edges in the drawing are usually shorter than for $k=4$, and their length is quite uniform. When $k$ increases, many edges become longer, due to a broader influence of the repulsive forces.

Figure~\ref{fi:add32} shows an example of drawings computed by \algo for the different values of $k$. One can see that the layout is progressively improved while $k$ increases. The figure also depicts a drawing of the same graph computed by OGDF-FR. Figure~\ref{fi:PGPGiantCompo} shows layouts of the graphs grund and pGp-giantcompo computed by \algo with $k=5$ and by OGDF-FR.

Concerning the synthetic graphs, we have a similar behavior of the quality metrics (see Tables~\ref{ta:synther-results-cr} and~\ref{ta:synthba-results-cr}). In particular, the number of crossings in the drawing of \algo is significantly reduced while passing from $k=2$ to $k=4$, although the final value is closer to that of the drawings computed by OGDF-FR. Also the similarity values usually improve while $k$ is increased and, for $k=4$, the values of \algo on the \synther graphs are always better than those of OGDF-FR.    

\begin{table}
\centering
\renewcommand{\arraystretch}{1.3}
\footnotesize
\begin{tabular}{| l | r | r | r | r | r | r | }
    \hline
    \rowcolor{DGray}
     & \multicolumn{3}{c|}{\algo - $k=2$} & \multicolumn{3}{c|}{\algo - $k=3$} \\ \cline{2-7}
	\rowcolor{Gray}    
    $|E|$ & CRE & ELD & SIM & CRE & ELD & SIM \\\hline
    10,000 & 374.43 & 9.36 & 0.37 & 304.62 & 17.64 & 0.72 \\
	50,000 & 1,536.98 & 8.77 & 0.56 & 1,247.99 & 16.12 & 0.68 \\
	100,000 & 3,470.94 & 9.24 & 0.56 & 2,833.03 & 17.43 & 0.74 \\
	1,000,000 & 29,510.85 & 8.68 & 0.87 & 24,163.59 & 15.98 & 0.93 \\ 
	1,500,000 & 49,665.11 & 9.09 & 0.61 & 40,809.52 & 17.06 & 0.93 \\
	2,000,000 & 78,383.53 & 9.12 & 0.94 & 55,913.71 & 17.18 & 1.00 \\

    \hline
    \rowcolor{DGray}
     & \multicolumn{3}{c|}{\algo - $k=4$} & \multicolumn{3}{c|}{OGDF FR} \\ \cline{2-7}
	\rowcolor{Gray}     
     $|E|$ & CRE & ELD & SIM & CRE & ELD & SIM  \\\hline
	10,000 & 267.89 & 31.08 & 1.00 & 266.77 & 51.55 & 0.49 \\
	50,000 & 1,100.38 & 28.44 & 1.00 & 1,066.38 & 95.94 & 0.43 \\
	100,000 & 2,518.16 & 31.83 & 1.00 & 2,445.86 & 114.86 & 0.44 \\
	1,000,000 & 21,376.95 & 28.13 & 1.00 & $*$ & $*$ & $*$ \\
	1,500,000 & 32,748.11 & 28.66 & 1.00 & $*$ & $*$ & $*$ \\
	2,000,000 & $*$ & $*$ & $*$ & $*$ & $*$ & $*$ \\
    \hline
  \end{tabular}
  \caption{\small Average number of crossings per edge (CRE), edge length standard deviation (ELD), and similarity (SIM) for the \synther graphs. For each graph, the similarity values are normalized between $0$ and $1$, so that $1$ corresponds to the best value.}\label{ta:synther-results-cr}
\end{table}

\begin{table}
\centering
\renewcommand{\arraystretch}{1.3}
\footnotesize
\begin{tabular}{| l | r | r | r | r | r | r | }
    \hline
    \rowcolor{DGray}
     & \multicolumn{3}{c|}{\algo - $k=2$} & \multicolumn{3}{c|}{\algo - $k=3$} \\ \cline{2-7}
	\rowcolor{Gray}    
    $|E|$ & CRE & ELD & SIM & CRE & ELD & SIM \\\hline
	$10,000$ & 437.86 & 14.57 & 0.03 & 372.61 & 30.58 & 0.17 \\
	$50,000$ & 1,634.32  & 16.36 & 0.00 & 1,328.27 & 37.84 & 0.24 \\
	$100,000$ & 3,250.08 & 17.47 & 0.00 & 2,649.02 & 42.15 & 0.08 \\
	$1,000,000$ & 40,811.54 & 19.47 & 1.00 & $*$ & $*$ & $*$ \\
	$1,500,000$ & 50,803.80 & 19.24 & 1.00 & $*$ & $*$ & $*$ \\
	$2,000,000$ & 78,383.53 & 20.10 & 1.00 & $*$ & $*$ & $*$ \\
    \hline
    \rowcolor{DGray}
     & \multicolumn{3}{c|}{\algo - $k=4$} & \multicolumn{3}{c|}{OGDF FR} \\ \cline{2-7}
    \rowcolor{Gray}     
     $|E|$ & CRE & ELD & SIM & CRE & ELD & SIM  \\\hline
	$10,000$ & 336.41 & 44.74 & 0.69 & 345.47 & 51.77 & 1.00 \\
	$50,000$ & 1,227.78 & 66.34 & 0.71 & 1,235.16 & 88.32 & 1.00 \\
	$100,000$ & 1,634.81  & 74.93 & 0.75 & 1,967.13 & 111.51 & 1.00 \\
	$1,000,000$ & $*$ & $*$ & $*$ & $*$ & $*$ & $*$ \\
	$1,500,000$ & $*$ & $*$ & $*$ & $*$ & $*$ & $*$ \\
	$2,000,000$ & $*$ & $*$ & $*$ & $*$ & $*$ & $*$  \\
    \hline
  \end{tabular}
  \caption{\small Average number of crossings per edge (CRE), edge length standard deviation (ELD), and similarity (SIM) for the \synthba graphs. For each graph, the similarity values are normalized between $0$ and $1$, so that $1$ corresponds to the best value.}\label{ta:synthba-results-cr}
\end{table}

\smallskip
Overall, the results about the quality metrics are still in favor of hypothesis \texttt{H2}.
We finally observe that also hypothesis \texttt{H3} seems to be confirmed by the experimental results. Indeed, looking at the real instances (for which we have different values of the graph diameter), the running time is most often negatively affected by small values of the diameter when $k$ increases. For the cluster with $10$ machines, the graphs for which we had the greatest increment of the running time while passing from $k=2$ to $k=4$ are p2p-Gnutella04, ca-CondMat, and p2p-Gnutella31, i.e., those with the smallest diameter. For the same graphs we observed the highest improvement in the number of crossings, together with graph grund, whose diameter is also relatively small. On the opposite, for the graph roadNet-PA, whose diameter is very large, the running time did not increase too much from $k=2$ to $k=4$, and the number of crossings improved by only $0.1\%$.

\section{An Application to Visual Cluster Detection}\label{se:application}
Big graphs from real-world applications are often small-world networks, locally dense, with an intrinsic community structure. For these graphs, most force-directed algorithms, included those based on the classical FR energy model, tend to produce cluttered drawings with hairball effects, which are not suitable to get detailed information about nodes and their connectivity. Instead, there are force-directed layout algorithms specifically conceived to give an overview of the graph structure in terms of its clusters. Among them, the LinLog energy model proposed by Noack is one of the most popular~\cite{DBLP:journals/jgaa/Noack07}. 

We applied our distributed force-directed technique to visual cluster detection in big graphs. Namely, we experimented the following approach: $(i)$ Compute a drawing of the input graph using \algo with the LinLog energy model in place of the classical FR model\footnote{Observe that the force function deriving from the LinLog energy model can be obtained as a particular tuning of the parameters in the FR force function.}; $(ii)$ Apply a $K$-means algorithm~\cite{k-means} to the set of points corresponding to the node positions (disregarding the edges), in order to compute a suitable set $K$ of node clusters. We automatically determine $K$ using a local search in a neighborhood of the initial value $K_0=\sqrt{n/2}$ (where $n$ is the number of nodes), and taking the value for which the corresponding clustering is the best one according to the Calinski-Harabasz qualitative index~\cite{dubes.93}. $(iii)$ Each cluster is then assigned a different color, which is used to display its nodes.

\begin{figure}
\centering
\subfigure[add32]{\includegraphics[width=0.45\columnwidth]{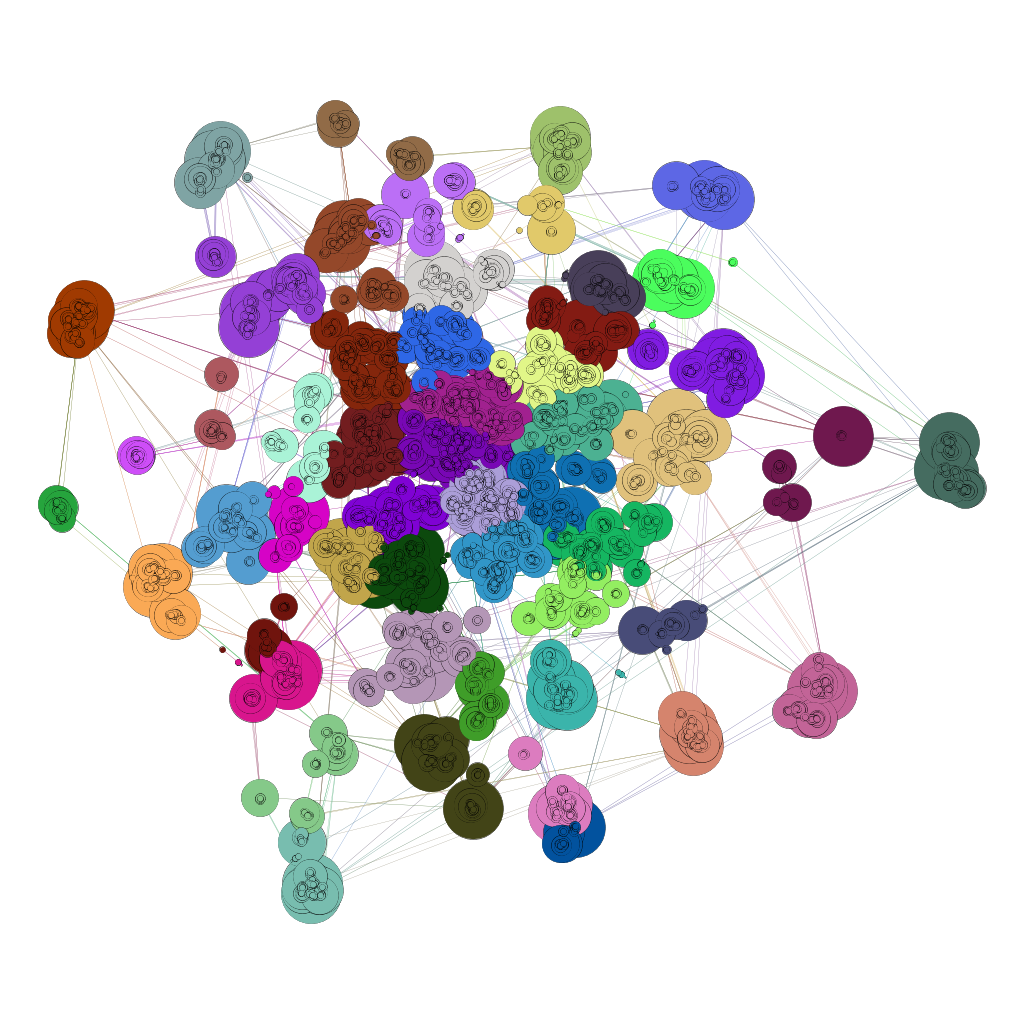}\label{fi:linlog-add32_K5}}\hfil
\subfigure[ca-GrQc]{\includegraphics[width=0.45\columnwidth]{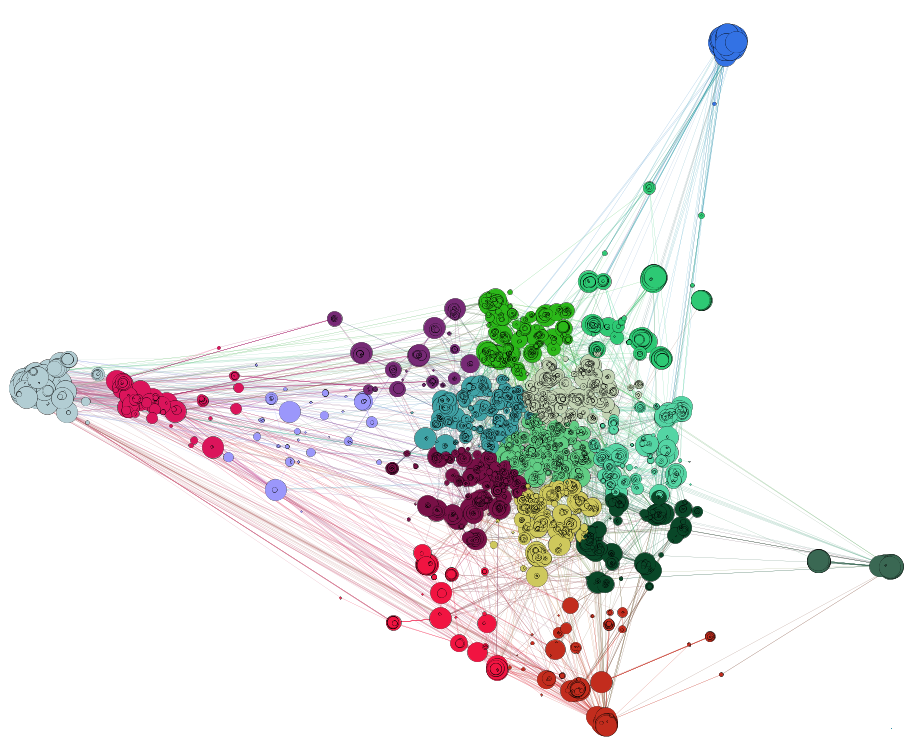}\label{fi:linlog-CA_GrQC_K4}}
\subfigure[pGp-giantcompo]{\includegraphics[width=0.45\columnwidth]{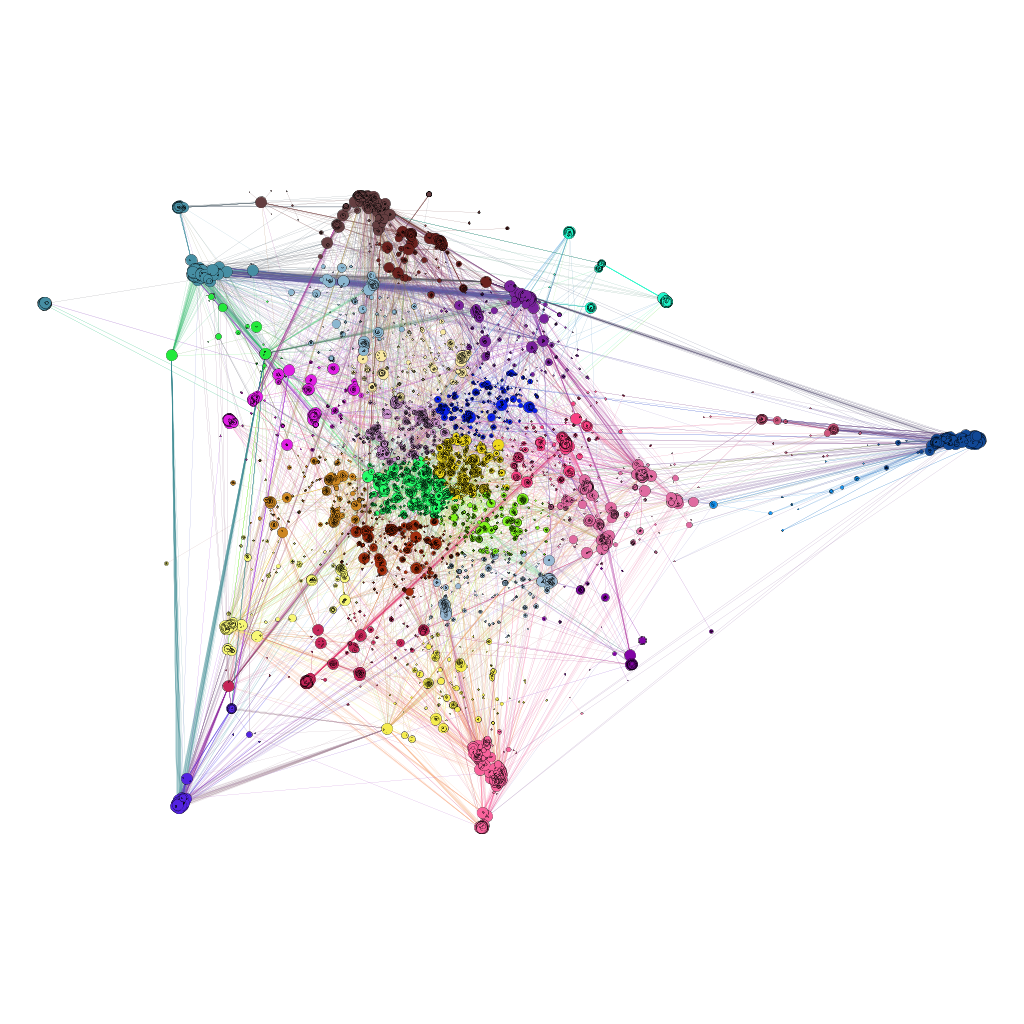}\label{fi:linlog-PGPGiantCompo_K5}}
\subfigure[p2p-Gnutella31]{\includegraphics[width=0.45\columnwidth]{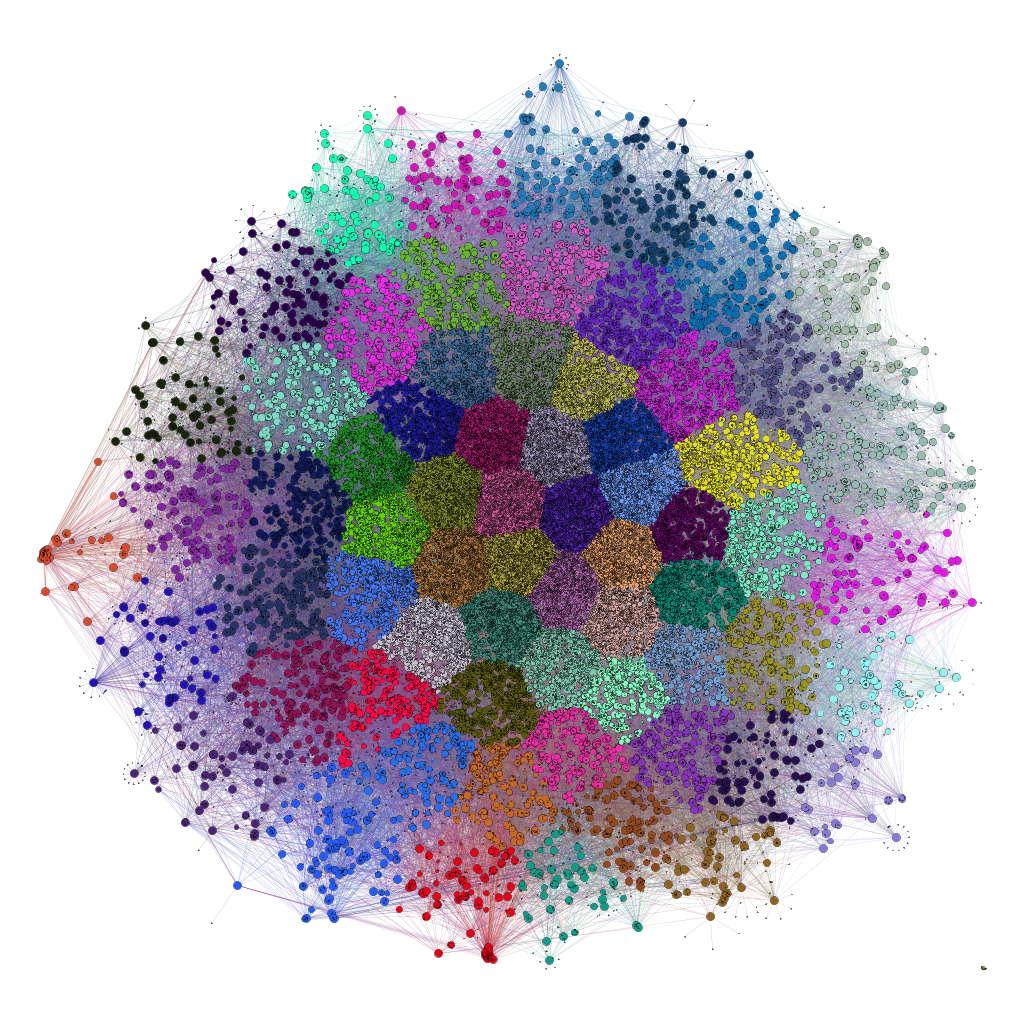}\label{fi:linlog-Gnutella31_K4}}\hfil
\caption{Examples of drawings computed by \algo, using the LinLog energy model.}\label{fi:linlog}
\end{figure}

\begin{figure}
\centering
\includegraphics[width=0.7\columnwidth]{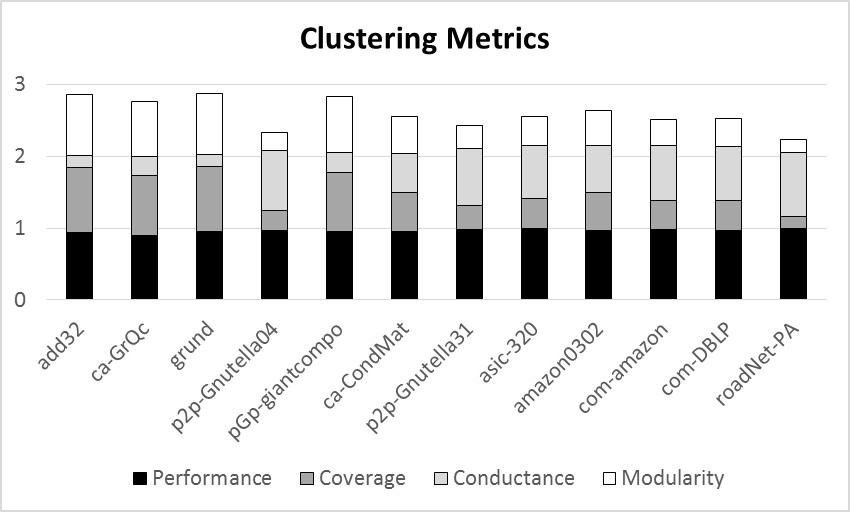}
\caption{\small Chart summarizing different quality metrics of the visual clustering computed by our approach on the \real graphs.}\label{fi:clusteringmetrics}
\end{figure}

Figure~\ref{fi:linlog} shows examples of visualizations produced with this approach for some of the graphs in the \real benchmark. In order to understand whether the visual clusters perceived by the user correspond to good clusters according to the connectivity of the graph, we measured different widely-used graph clustering quality metrics, namely \emph{performance}, \emph{coverage}, \emph{conductance}, and \emph{modularity} (see, e.g.,~\cite{DBLP:conf/pkdd/AlmeidaNMZ11,DBLP:journals/tkde/BrandesDGGHNW08,DBLP:journals/jea/BrandesGW07}). The chart in Figure~\ref{fi:clusteringmetrics} summarizes the values of these metrics for the \real graphs, on which \algo ran with the LinLog energy model, with the maximum $k$ for which it succeeded to compute a drawing. All values are normalized in the interval $[0,1]$, where $1$ is the optimum value.

The performance of the clustering is always quite good (close to the optimal value). The conductance ($0.57$ on average) on the smaller graphs is often low, but it tends to increase for the larger graphs. The values of coverage ($0.55$ on average) and modularity ($0.51$ on average) are more oscillating, but they are still relatively high for several instances. 
 
\section{Conclusions and Future Research}\label{se:future}
We presented \algo, the first distributed force-directed algorithm running on the Giraph framework, which is based on a vertex centric paradigm. Compared to previous parallel and distributed graph layout algorithms, it appears to be faster and more scalabale to large graphs. We showed that the algorithm can successfully run on an inexpensive cloud computing platform: layouts of graphs with one million edges can be computed in few minutes with a cost of few dollars. The code of our implementation is made available over the web. 

In the near future, we plan to develop a distributed multi-level force-directed algorithm, still based on the TLAV paradigm. The design of such an algorithm is challenging, due to the intrinsic difficulty of efficiently computing the hierarchy required by a multi-level approach in a distributed manner, as also observed in~\cite{DBLP:conf/iv/AntoineD15}. We believe that our TLAV approach can be conveniently exploited to achieve this goal, and that \algo can be used as a single-level force-directed algorithm to refine the layouts generated at the different levels of the hierarchy. We recall that multi-level force-directed algorithms run much faster than single-level ones, and often produce qualitatively better layouts (see~\cite{Handbook-kob} for a high-level description and references about how multi-level force-directed algorithms work).     

\bibliographystyle{abbrv}
{\small \bibliography{distributed}}

\end{document}